\documentclass[a4paper,12pt]{article}
\usepackage{amssymb}
\usepackage{amsmath}
\usepackage{epsfig}
\usepackage{caption}
\usepackage{subfigure}
\usepackage{graphicx}
\usepackage{tikz}
\usepackage{latexsym}
\usepackage{rotating}
\usepackage{titlesec}
\newcommand{\squeezeup}{\vspace{-10.5mm}}
\usetikzlibrary{matrix}

\title{The Method of Hirota  Bilinearization}

\author{Metin G\"{u}rses \thanks{gurses@fen.bilkent.edu.tr}\\
{\small Department of Mathematics, Faculty of Science}\\
{\small Bilkent University, 06800 Ankara - Turkey}\\
Asl{\i} Pekcan \thanks{aslipekcan@hacettepe.edu.tr} \\
{\small Department of Mathematics, Faculty of Science} \\
{\small Hacettepe University, 06800 Ankara - Turkey}
}

\date{\nonumber}
\begin{document}
\maketitle
\date{\nonumber}
\newtheorem{thm}{Theorem}[section]
\newtheorem{Le}{Lemma}[section]
\newtheorem{defi}{Definition}[section]
\newtheorem{ex}{Example}[section]
\newtheorem{pro}{Proposition}[section]
\baselineskip 17pt

\numberwithin{equation}{section}

\begin{abstract}
Bilinearization of a given nonlinear partial differential equation is very important not only to find soliton solutions but also to obtain other solutions such as  the complexitons, positons, negatons, and lump solutions. In this work we study the bilinearization of nonlinear partial differential equations in $(2+1)$-dimensions. We write the most general sixth order Hirota bilinear form in $(2+1)$-dimensions and give the associated nonlinear partial differential equations for each monomial of the product of the Hirota operators $D_{x}$, $D_{y}$, and $D_{t}$.
The nonlinear partial differential equations corresponding to the sixth order Hirota bilinear equations are in general nonlocal. Among all these we give the most general sixth order Hirota bilinear equation whose nonlinear partial differential equation is local which contains 12 arbitrary constants. Some special cases of this equation are the KdV, KP, KP-fifth order KdV, and Ma-Hua equations. We also obtain a nonlocal nonlinear partial differential equation whose Hirota form contains all possible triple products of $D_{x}$, $D_{y}$, and $D_{t}$. We give one- and two-soliton solutions, lump solutions with one, two, and three functions, and hybrid solutions of local and nonlocal $(2+1)$-dimensional equations. We proposed also solutions of these equations depending on dynamical variables.
\end{abstract}


\section{Introduction}
Integrability of nonlinear partial differential equations is one of the major research area in applied mathematics and mathematical physics.
There are various ways of attacking to such problems. The standard way is to search for Lax pairs associated to a given system of nonlinear partial differential equations. Another and equivalent approach is to obtain a recursion operator where one can write all the hierarchy of higher symmetries in a compact form. There are also other important approaches such as the Hirota bilinear formalism \cite{hir}-\cite{hit5}. These approaches are particularly  more effective and practical for non-evolutionary equations. In particular we observed that the Hirota method played a very important role in finding soliton solutions of nonlocal integrable nonlinear partial differential equations \cite{gur1}-\cite{gur6}. However the integrability of nonlinear partial differential equations by Hirota  approach may not imply the Lax integrability and existence of recursion operators. In spite of this fact there is an increasing interest in Hirota integrability in the last decade. The definition of Hirota integrability was given by Hietarinta \cite{hit}. If
any number $N$ of one-soliton solutions of an equation can be combined into $N$-soliton solution consisting of exponential functions then the equation is Hirota integrable.
The first attempt to search for integrable nonlinear partial differential equations admitting Hirota bilinear form is due to Hietarinta \cite{hit}. He proposed the following Hirota bilinear form
\begin{equation}
\left( D_x^4-D_x\,D_t^3+a D_x^2+b D_x\,D_t+c D_t^2 \right) \{f \cdot f\}=0, \label{hit}
\end{equation}
where $a, b,$ and $c$ are constants. Hietarinta showed that this equation has four-soliton solutions and passes the Painlev\'{e} test.
Hietarinta's bilinear equation in (\ref{hit}) corresponds to the following nonlinear (nonlocal) partial differential equation in $(1+1)$-dimension
\begin{equation}
u_{xxxx}+6 u_{x} u_{xx}-u_{tttx}-3 u_{t} u_{tt}-3 u_{xt}\, (D^{-1}\, u_{tt})+a u_{xx}+b u_{xt}+c u_{tt}=0,
\end{equation}
which can be made local when we let $u=v_{x}$.

In this work we shall continue on the Hietarinta's work further by increasing the powers of the operators $D_{x}, D_{y}$, and $D_{t}$ in $(2+1)$-dimensions.
In the last decade we observe some interesting works on extending the works of Hietarinta. In particular Ma and his colleagues produced some examples of $(2+1)$-dimensional Hirota integrable partial differential equations \cite{Ma-0}-\cite{Ma-6}, \cite{Meng}, \cite{Lin}. The main idea in all these works is to  obtain the soliton solutions of the associated nonlinear partial differential equations for a given Hirota bilinear form. Another important reason for studying Hirota bilinearization is to obtain not only soliton solutions by the Hirota method but also other kind of solutions such as the complexitons, positons, negatons, lump, and hybrid solutions of the nonlinear partial differential equations whose bilinearization is possible.

Let $(x,y,t)$ denote the independent variables and $u=a_0 (\ln f)_{x}$ be the dependent variable. Here $a_{0}$ is a constant. Then the most general Hirota bilinear form, having only binary products of order $2n$ is given by
\begin{equation}
\sum_{k=0}^{2n} \left[\alpha_{k}\,D^{k}_{x}\,D^{2n-k}_{t}+\beta_{k}\,D^{k}_{x}\,D^{2n-k}_{y}+\gamma_{k}\,D^{k}_{y}\,D^{2n-k}_{t} \right] \{f \cdot f\}=0,
\end{equation}
where $\alpha_{k}, \beta_{k}, \gamma_{k}$ are constants to be adjusted to obtain new integrable equations in $(2+1)$-dimensions. Superposition of the above forms may be given as
\begin{equation}
\sum_{n=1}^{N}\,\sum_{k=0}^{2n}  \left[\alpha_{kn}\,D^{k}_{x}\,D^{2n-k}_{t}+\beta_{kn}\,D^{k}_{x}\,D^{2n-k}_{y}+\gamma_{kn}\,D^{k}_{y}\,D^{2n-k}_{t} \right]\,\{f \cdot f\}=0,
\end{equation}
where $N$ is a natural number and $\alpha_{kn}, \beta_{kn}$, and $\gamma_{kn}$ are constants. In the above forms we have only binary products of $D_x, D_y$, and $D_t$. We have further generalization of these bilinear forms by including the triple products of these operators as well:
\begin{equation}
\left(\sum_{m=1}\, \sum_{n=1} \, C_{nm}\, D_{t}^m\, D_{x}^n\, D_{y}^{N-m-n} \right) \{f \cdot f\}=0,
\end{equation}
where $C_{nm}$'s are constants. Any bilinear form of degree $N$ in $(2+1)$-dimensions must contain both binary and triple products of the operators $D_x, D_y$, and $D_t$.

In  Appendix A we give all bilinear forms with four and six orders in $(2+1)$-dimensions.
One of the main purpose of this work  is to obtain nonlinear partial differential equations corresponding to the above bilinear forms by letting $u=2\, (\ln f)_{x}$. When $N=3$ we have the following most general sixth order Hirota bilinear form in $(2+1)$-dimensions
\begin{equation}
F(D_x, D_y, D_t)\, \{f \cdot f\}=0, \label{genhir}
\end{equation}
where
\begin{align}
F(D_x, D_y, D_t)&=\sum_{k=0}^{2}  \left[\alpha_{k1}\,D^{k}_{x}\,D^{2-k}_{t}+\beta_{k1}\,D^{k}_{x}\,D^{2-k}_{y}+\gamma_{k1}\,D^{k}_{y}\,D^{2-k}_{t} \right] \nonumber\\
&+\sum_{k=0}^{4}  \left[\alpha_{k2}\,D^{k}_{x}\,D^{4-k}_{t}+\beta_{k2}\,D^{k}_{x}\,D^{4-k}_{y}+\gamma_{k2}\,D^{k}_{y}\,D^{4-k}_{t} \right]\nonumber\\
&+\sum_{k=0}^{6}  \left[\alpha_{k3}\,D^{k}_{x}\,D^{6-k}_{t}+\beta_{k3}\,D^{k}_{x}\,D^{6-k}_{y}+\gamma_{k3}\,D^{k}_{y}\,D^{6-k}_{t} \right] \nonumber\\
&+\sum_{m=1}^{3}\, \sum_{n=1}^{3-m}\, B_{nm}\, D_{t}^m \, D_{x}^n\, D^{4-m-n}_{y}+\sum_{m=1}^{5}\, \sum_{n=1}^{5-m}\, C_{nm}\, D_{t}^m \, D_{x}^n\, D^{6-m-n}_{y}, \label{genhir1}
\end{align}
where $\alpha_{ij}$, $\beta_{ij}$, $\gamma_{ij}$, $B_{ij}$, and $C_{ij}$ are constants. Now we list the important results in this work.

\vspace{0.2cm}
\noindent
{\bf 1. The most general bilinear form of sixth degree and the nonlinear partial differential equation associated with it}:
The above equations (\ref{genhir}) and (\ref{genhir1}) give  the most general Hirota bilinear equation of sixth order in $(2+1)$-dimensions. It is straightforward to find the corresponding nonlinear differential equations via $u=2 (\ln f)_{x}$ by using the list of monomials of Hirota bilinear forms in Appendix A. It is of course  not so practical to obtain the above partial differential equation in its general form. Hence we will focus on some special cases which cover the most of the well-known nonlinear differential equations of third and fifth order in $(2+1)$-dimensions.

How to use the Appendix A is as follows: Let the nonlinear differential equation be $E(u,u_{t},u_{x}, u_{y}, \cdots)=0$ then the equation corresponding to the Hirota bilinear equation (\ref{genhir}) is obtained through
\begin{equation}
\left(\frac{F(D_x, D_y, D_t)\, \{f \cdot f\}}{f^2} \right)_{x}=E(u,u_{t},u_{x}, u_{y}, \cdots), \label{genhir2}
\end{equation}
by letting $u=2\,(\ln f)_{x}$.

By using some ansatzes on the function $f$ we can obtain different type of solutions such as  solitons, complexitons, positons, negatons, and lumps.

\vspace{0.2cm}
\noindent
{\bf 2.  Some special cases of the equation (\ref{genhir2})}:
We first obtain the most  general sixth order local nonlinear partial differential equation in $(2+1)$-dimensions obtainable from (\ref{genhir})-(\ref{genhir1}) and then a nonlocal partial differential equation in $(2+1)$-dimensions obtained by including triple bilinear forms.

\vspace{0.2cm}
\noindent
{\bf 3.  Soliton, lump, hybrid solutions and solutions depending on dynamical variables}:
We obtain soliton, lump, and hybrid solutions of the special equations we obtain. All the solitonic solutions we obtain look similar to those soliton solutions of integrable nonlinear partial differential equations \cite{gur5}, \cite{gur6}, \cite{gur7} in $(2+1)$-dimensions. We obtain first one- and two-soliton solutions of these equations. We observe that for three-soliton solutions the parameters of the equations should satisfy certain conditions. We then obtain lump solutions with one, two, and three functions.  The lump solutions are rational functional solutions localized in all directions in the space which were first discovered by Manakov et al. \cite{Manakov}. We are able to find the mixture of soliton and the lump solutions which are named as hybrid solutions. In this work we also consider solutions of our special equations depending on some dynamical variables in three dimensions. As far as we know such solutions are new in literature.

We have three Appendicies at the end. Appendix A contains all monomials of Hirota bilinear form and the corresponding expressions of the field variable $u(x,y,t)$. Appendix B and Appendix C contain long expressions obtained to find the lump and hybrid solutions of our special equations (\ref{local1}) and (\ref{non1}).

\section{Some special equations}

Nonlinear partial differential equations in $(2+1)$-dimensions obtainable from (\ref{genhir}) and (\ref{genhir1}) are mostly nonlocal as we can see from the list given in the Appendix A. The differential equation associated to the general bilinear form (\ref{genhir}) and (\ref{genhir1}) is quite lengthy and complicated. Instead of studying this general equation we prefer to consider some special cases which cover most of the integrable differential equations.

\subsection{The most general local equation}

 As the first example
we now wish to give the most general Hirota bilinear equation of sixth degree which leads to a local differential equation in $(2+1)$-dimensions.
\begin{align}\label{local1Hir}
\Big(a_{0}\, D_{t}^2+&a_{1}\, D_{x}^2+a_{2}\, D_{y}^2+a_{3}\,D_{t}\, D_{x}+a_{4}\, D_{t}\, D_{y}+a_{5}\, D_{x}\,D_{y}  \nonumber\\
&+b_{0}\,D_{y}\,D_{x}^3+b_{1}\,D_{t}\,D_{x}^3+b_{2}\, D_{x}^4+c_{0}\,D_{x}^6+c_{1}\, D_{t}\,D_{x}^5+c_{2}\,D_{y}\,D_{x}^5 \Big)\,\{f \cdot f\}=0,
\end{align}
where $a_{0},a_{1},a_{2},a_{3},a_{4},a_{5}$, $b_{0},b_{1},b_{2}$, $c_{0},c_{1},c_{2}$ are constants. This bilinear equation, through $u=2\,(\ln f)_{x}$, leads to the following nonlinear partial differential equation.
\begin{eqnarray}
&&a_{0}\,u_{tt}+a_{1}\,u_{xx}+a_{2}\,u_{yy}+a_{3}\,u_{tx}+a_{4}\,u_{ty}+a_{5}\,u_{xy}+b_{0}\,(u_{yxx}+3 u_{x} u_{y})_{x} \nonumber\\
&&+b_{1}\,(u_{txx}+3 u_{t} u_{x})_{x}+b_{2}\,(u_{xxx}+3 u_{x}^2)_{x} +c_{0}\,(u_{xxxxx}+15 u_{x} u_{xxx}+ 15 u_{x}^3)_{x} \nonumber \\
&&+c_{1}\,(u_{txxxx}+10 u_{x} u_{xxt}+5 u_{t} u_{xxx}+15u_x^2u_t)_{x}
\nonumber \\
&&+c_{2}\,(u_{yxxxx}+10 u_{x} u_{xxy}+5 u_{y} u_{xx}+15u_x^2u_y)_{x}=0.\label{local1}
\end{eqnarray}
We have the following special cases: $a_{3} \ne 0$, $b_{2} \ne 0$, (KdV equation); $a_3\ne 0$, $c_0\ne 0$, (Sawada-Kotera equation); $a_{2} \ne 0$, $a_{3} \ne 0$,  $b_{2} \ne 0$, (KP equation; $a_0\ne 0$, $a_1\ne 0$, $b_2\ne 0$, (Boussinesq equation); $a_0\ne 0$, $a_1\ne 0$, $b_2\ne 0$, $c_0\ne 0$, (Higher order Boussinesq equation \cite{BoRen}); $a_{3} \ne 0$, $b_{2} \ne 0$, $c_{0} \ne 0$,  (KP+Fifth order KdV equation); $a_0\ne 0$, $b_1\ne 0$, $c_0\ne 0$, (KdV(6) equation \cite{karasu});
$a_{1} \ne 0$, $a_{2} \ne 0$, $a_{4} \ne 0$, $b_{0} \ne 0$, (Ma-Hua equation \cite{Ma-2}); $a_{1} \ne0$, $a_{4} \ne 0$, $b_{1} \ne 0$, (Hirota-Satsuma-Ito equation \cite{hit});
$a_0\ne 0$, $b_1\ne 0$, $c_0\ne 0$, (6th order Ramani equation \cite{ramani}); $a_0\ne 0$, $a_5\ne 0$, $b_1\ne 0$, $c_0\ne 0$, (Sawada-Kotera-Ramani equation);
$a_j\ne 0$, $j=1,2,3,4,5$, $b_1\ne 0$, (Generalized Hirota-Satsuma-Ito equation \cite{MLK}, \cite{HTW}); $a_2\ne 0$,  $a_3\ne 0$, $b_0\ne 0$, $c_0\ne 0$, (BKP equation \cite{Date1},  \cite{JimboMiwa}), $a_1 \ne 0$, $a_2 \ne 0$, $a_3\ne 0$, $b_1\ne 0$, (KP-Benjamin-Bona-Mahony equation);
$a_3 \ne 0$, $b_0 \ne 0$, $b_2 \ne 0$, (Generalized Bogoyavlensky-Konopelchenko equation \cite{Calogero}, \cite{Bogo});
 $a_1 \ne 0$, $a_5 \ne 0$, $b_0 \ne 0$, $c_0 \ne 0$, (Generalized BKP equation \cite{Ma-0}); $a_j\ne 0$, $j=0,1,2,3,4,5$, $b_2\ne 0$, (Generalized KP equation \cite{Ma-0}).

  We obtain one- and two-soliton solutions of these equations in Section 4, lump solutions with one, two, and three functions in Section 5, hybrid  solutions in Section 6, and solutions depending on some dynamical variables in three dimensions in Section 7.

\subsection{A nonlocal equation}

The general sixth order nonlinear partial differential equations associated to the Hirota bilinear equation (\ref{genhir}) are mostly nonlocal. Although these equations are nonlocal we can solve them obtaining one-, two-, and three-soliton solutions by using the Hirota method. We can consider the following particular case of (\ref{genhir1}),
\begin{equation}\label{non1Hir}
\left(a_{0}\,D_{t}\, D_{x}+a_{1}\,D_{y}^2 +a_{2}\, D_{x}^4+ b_{0}\,D_{t}\,D_x^2\,D_y +b_{1}\,D_{t}^2\,D_x\,D_y \, + b_{2}\,D_{t}\,D_x\,D_y^2 \right) \{f \cdot f\}=0,
\end{equation}
where $a_{0},a_{1},a_{2}$, $b_{0},b_{1},b_{2}$ are constants. The above bilinear equation, through $u=2\,(\ln f)_{x}$, leads to the following nonlocal nonlinear partial differential equation:
\begin{eqnarray}
&&a_{0}\, u_{xt}+a_{1}\,u_{yy}+a_{2} (u_{xxx}+3 u_{x}^2)_{x}+b_{0}\,(u_{xxyt}+u_xu_{yt}+2u_yu_{xt}+2u_tu_{xy}+u_{xx}\,D^{-1}\,u_{yt})\nonumber\\
&&+b_{1}\,(u_{xytt}+u_yu_{tt}+2u_tu_{yt}+u_{xy}\,D^{-1}\,u_{tt}+2u_{xt}\,D^{-1}\,u_{yt}) \nonumber\\
&&+b_{2}\,(u_{xtyy}+u_tu_{yy}+2u_yu_{yt}+u_{xt}\,D^{-1}\,u_{yy}+2u_{xy}\, D^{-1}\,u_{yt})=0, \label{non1}
\end{eqnarray}
where $D^{-1}\, f=\int^{x}\, f\, dx'$ for any $f$. The above equation can be made local by defining new dependent variable $u=v_{x}$. This nonlocal equation (\ref{non1}) is a generalization of the KP equation where the corresponding Hirota bilinear form is a fourth order equation. We obtain one- and two-soliton solutions of this equation in Section 4, lump solutions in Section 5, and hybrid solutions in Section 6.

\section{Solitonic Solutions}

\subsection{Soliton solutions of (\ref{local1})}

\vspace{0.cm}
\noindent
{\bf 1.\, One-soliton solutions of (\ref{local1})}

\vspace{0.cm}
\noindent
To obtain one-soliton solutions of (\ref{local1}) we take $f=1+\varepsilon f_1$ where $f_1=e^{k_1x+\omega_1t+l_1y+\alpha_1}$ for
arbitrary constants $k_1, \omega_1, l_1, \alpha_1$ and insert it into (\ref{local1Hir}). Analyzing the coefficients of the powers of $\varepsilon$ gives
the dispersion relation as
\begin{align}
l_1=&\frac{1}{2a_2}(-b_0k_1^3-a_4\omega_1-a_5k_1-c_2k_1^5\pm[b_0^2k_1^6+2b_0a_4k_1^3\omega_1+2b_0a_5k_1^4+2b_0c_2k_1^8+a_4^2\omega_1^2\nonumber\\
&+2a_4a_5\omega_1k_1+2a_4c_2\omega_1k_1^5
+a_5^2k_1^2+2a_5c_2k_1^6+c_2^2k_1^{10}-4a_2a_0\omega_1^2
-4a_2a_1k_1^2\nonumber\\
&-4a_2c_1\omega_1k_1^5-4a_2a_3\omega_1k_1-4a_2b_2k_1^4-4a_2c_0k_1^6-4a_2b_1\omega_1k_1^3)]^{1/2}).
\end{align}
Letting $\varepsilon=1$ we obtain one-soliton solutions of the equation (\ref{local1}) as
\begin{equation}
u(x,y,t)=2(\ln(f(x,y,t)))_x=k_1e^{(k_1x+\omega_1t+l_1y+\alpha_1)/2}\mathrm{sech}((k_1x+\omega_1t+l_1y+\alpha_1)/2).
\end{equation}

%\newpage
\noindent
{\bf 2.\, Two-soliton solutions of (\ref{local1})}

\vspace{0.cm}
\noindent
Let $f=1+\varepsilon f_1+\varepsilon^2 f_2$ where $f_1=e^{\theta_1}+e^{\theta_2}$ for $\theta_j=k_jx+\omega_jt+l_jy+\alpha_j$, $j=1, 2$.
Inserting $f$ into the Hirota bilinear form (\ref{local1Hir}) and making the coefficients of the powers of $\varepsilon$ zero yield
the dispersion relations
\begin{align}\label{displocal12SS}
l_j=&\frac{1}{2a_2}(-b_0k_j^3-a_4\omega_j-a_5k_j-c_2k_j^5\pm [b_0^2k_j^6+2b_0a_4k_j^3\omega_j+2b_0a_5k_j^4+2b_0c_2k_j^8+a_4^2\omega_j^2\nonumber\\
&+2a_4a_5\omega_jk_j+2a_4c_2\omega_jk_j^5
+a_5^2k_j^2+2a_5c_2k_j^6+c_2^2k_j^{10}-4a_2a_0\omega_j^2
-4a_2a_1k_j^2\nonumber\\
&-4a_2c_1\omega_jk_j^5-4a_2a_3\omega_jk_j-4a_2b_2k_j^4-4a_2c_0k_j^6-4a_2b_1\omega_jk_j^3]^{1/2})
\end{align}
for $j=1, 2$. The coefficient of $\varepsilon^2$ gives $f_2=A_{12}e^{\theta_1+\theta_2}$ for $A_{12}=-\frac{P_1(p_1-p_2)}{P_1(p_1+p_2)}$, where
\begin{align}
&P_1(p_1-p_2)=a_0(\omega_1-\omega_2)^2+a_1(k_1-k_2)^2+a_2(l_1-l_2)^2+a_3(\omega_1-\omega_2)(k_1-k_2)\nonumber\\
&+a_4(\omega_1-\omega_2)(l_1-l_2)+a_5(k_1-k_2)(l_1-l_2)+b_0(l_1-l_2)(k_1-k_2)^3\nonumber\\
&+b_1(\omega_1-\omega_2)(k_1-k_2)^3
+b_2(k_1-k_2)^4+c_0(k_1-k_2)^6+c_1(\omega_1-\omega_2)(k_1-k_2)^5\nonumber\\
&+c_2(l_1-l_2)(k_1-k_2)^5,\\
&P_1(p_1+p_2)=a_0(\omega_1+\omega_2)^2+a_1(k_1+k_2)^2+a_2(l_1+l_2)^2+a_3(\omega_1+\omega_2)(k_1+k_2)\nonumber\\
&+a_4(\omega_1+\omega_2)(l_1+l_2)+a_5(k_1+k_2)(l_1+l_2)+b_0(l_1+l_2)(k_1+k_2)^3\nonumber\\
&+b_1(\omega_1+\omega_2)(k_1+k_2)^3+b_2(k_1+k_2)^4+c_0(k_1+k_2)^6+c_1(\omega_1+\omega_2)(k_1+k_2)^5\nonumber\\
&+c_2(l_1+l_2)(k_1+k_2)^5.
\end{align}
Without loss of generality we take $\varepsilon=1$. Hence two-soliton solution of the equation (\ref{local1}) is
\begin{equation}
u(x,y,t)=2(\ln(f(x,y,t)))_x=\frac{2[k_1e^{\theta_1}(1+A_{12}e^{\theta_2})+k_2e^{\theta_2}(1+A_{12}e^{\theta_1})]}{1+e^{\theta_1}+e^{\theta_2}+A_{12}e^{\theta_1+\theta_2}},
\end{equation}
where $\theta_j=k_jx+\omega_jt+l_jy+\alpha_j$, $j=1, 2$, with the dispersion relations (\ref{displocal12SS}) satisfied.

\subsection{Soliton solutions of (\ref{non1})}

\vspace{0.cm}
\noindent
{\bf 1.\, One-soliton solutions of (\ref{non1})}

\vspace{0.cm}
\noindent
We insert $f=1+\varepsilon f_1$ where $f_1=e^{k_1x+\omega_1t+l_1y+\alpha_1}$ into (\ref{non1Hir}) to obtain one-soliton solutions of the equation (\ref{non1}). Here
$k_1, \omega_1, l_1, \alpha_1$ are arbitrary constants. Making the coefficients of the powers of $\varepsilon$ zero yields
the dispersion relation as
\begin{align}\label{dispnon1-1SS}
l_1=&\frac{1}{2(a_1+b_2\omega_1k_1)}(-b_0\omega_1k_1^2-b_1\omega_1^2k_1\pm [b_0^2\omega_1^2k_1^4+2b_0b_1\omega_1^3k_1^3+b_1^2\omega_1^4k_1^2
\nonumber\\
&-4a_1a_0\omega_1k_1-4a_1a_2k_1^4-4b_2a_0\omega_1^2k_1^2-4b_2a_2\omega_1k_1^5]^{1/2}).
\end{align}
We let $\varepsilon=1$ and obtain one-soliton solution of the equation (\ref{non1}) as
\begin{equation}
u(x,y,t)=2(\ln(f(x,y,t)))_x=k_1e^{(k_1x+\omega_1t+l_1y+\alpha_1)/2}\mathrm{sech}((k_1x+\omega_1t+l_1y+\alpha_1)/2),
\end{equation}
where the dispersion relation (\ref{dispnon1-1SS}) holds.\\

\vspace{0.cm}
\noindent
{\bf 2.\, Two-soliton solutions of (\ref{non1})}

\vspace{0.cm}
\noindent
To obtain two-soliton solutions of (\ref{non1}), we take $f=1+\varepsilon f_1+\varepsilon^2 f_2$ where $f_1=e^{\theta_1}+e^{\theta_2}$ for $\theta_j=k_jx+\omega_jt+l_jy+\alpha_j$, $j=1, 2$ and insert $f$ into the Hirota bilinear form (\ref{non1Hir}). Analyzing the coefficients of the powers of $\varepsilon$ we obtain
the dispersion relations as
\begin{align}\label{dispnon12SS}
l_j=&\frac{1}{2(a_1+b_2\omega_jk_j)}(-b_0\omega_jk_j^2-b_1\omega_j^2k_j\pm (b_0^2\omega_j^2k_j^4+2b_0b_1\omega_j^3k_j^3+b_1^2\omega_j^4k_j^2
\nonumber\\
&-4a_1a_0\omega_jk_j-4a_1a_2k_j^4-4b_2a_0\omega_j^2k_j^2-4b_2a_2\omega_jk_j^5)^{1/2})
\end{align}
for $j=1, 2$ and the function $f_2=A_{12}e^{\theta_1+\theta_2}$ for $A_{12}=-\frac{P_2(p_1-p_2)}{P_2(p_1+p_2)}$, where
\begin{align}
&P_2(p_1-p_2)=a_0(\omega_1-\omega_2)(k_1-k_2)+a_1(l_1-l_2)^2+a_2(k_1-k_2)^4\nonumber\\
&+b_0(\omega_1-\omega_2)(k_1-k_2)^2(l_1-l_2)+b_1(\omega_1-\omega_2)^2(k_1-k_2)(l_1-l_2)
\nonumber\\
&+b_2(\omega_1-\omega_2)(k_1-k_2)(l_1-l_2)^2,\\
&P_2(p_1+p_2)=a_0(\omega_1+\omega_2)(k_1+k_2)+a_1(l_1+l_2)^2+a_2(k_1+k_2)^4\nonumber\\
&+b_0(\omega_1+\omega_2)(k_1+k_2)^2(l_1+l_2)+b_1(\omega_1+\omega_2)^2(k_1+k_2)(l_1+l_2)\nonumber\\
&+b_2(\omega_1+\omega_2)(k_1+k_2)(l_1+l_2)^2.
\end{align}
Take $\varepsilon=1$. Hence two-soliton solutions of the equation (\ref{non1}) are given by
\begin{equation}
u(x,y,t)=2(\ln(f(x,y,t)))_x=\frac{2[k_1e^{\theta_1}(1+A_{12}e^{\theta_2})+k_2e^{\theta_2}(1+A_{12}e^{\theta_1})]}{1+e^{\theta_1}+e^{\theta_2}+A_{12}e^{\theta_1+\theta_2}},
\end{equation}
where $\theta_j=k_jx+\omega_jt+l_jy+\alpha_j$, $j=1, 2$, with the dispersion relations (\ref{dispnon12SS}) satisfied.

Solitonic solutions of (\ref{local1}) and (\ref{non1}) show exactly similar properties of the soliton solutions of the standard integrable equations. We shall investigate the three-soliton solutions of these equations in a later communication.

\subsection{Three-soliton solutions of (\ref{local1}) and (\ref{non1})}

We have studied three-soliton solutions of the equations (\ref{local1}) and (\ref{non1}). First of all there exists no three-soliton solutions with arbitrary values of parameters of the equations. This means that parameters $a_{0}, a_{1}, \cdots $, $b_{0},b_{1}, \cdots$ and $c_{0}, c_{1}, \cdots $ satisfy certain conditions. These constraints are quite lengthy and hence we leave the study on the three-soliton solutions of  (\ref{local1}) and (\ref{non1}) for a later communication.

\section{Lump solutions}

Lump solutions of bilinear equations are obtained by taking \cite{Ma-0}-\cite{Ma-6}, \cite{Feng}-\cite{BaiDeng},
\begin{equation}\label{Mlump}
f=\beta_0+\sum_{j=1}^M p_j,
\end{equation}
where $p_j=\beta_1^jx+\beta_2^jy+\beta_3^jt+\beta_4^j$, $\beta_{0}, \beta_{s}^j$, $s=1,2,3,4$, $j=1,\cdots, M$  are arbitrary constants.
Inserting (\ref{Mlump}) into the bilinear equations of $f$ we obtain some conditions on the constants $\beta_{0}, \beta_s^{j}$. The bilinear form of (\ref{local1})
is given by
\begin{align}\label{bilinearLOC1}
&a_0(ff_{tt}-f_t^2)+a_1(ff_{xx}-f_x^2)+a_2(ff_{yy}-f_y^2)+a_3(ff_{xt}-f_xf_t)+a_4(ff_{ty}-f_tf_y)\nonumber\\
&+a_5(ff_{xy}-f_xf_y)+b_0(ff_{xxxy}-3f_xf_{xxy}+3f_{xx}f_{xy}-f_yf_{xxx})\nonumber\\
&+b_1(ff_{xxxt}-3f_xf_{xxt}+3f_{xx}f_{xt}-f_tf_{xxx})+b_2(ff_{xxxx}-4f_xf_{xxx}+3f_{xx}^2)\nonumber\\
&+c_0(ff_{xxxxxx}-6f_xf_{xxxxx}+15f_{xx}f_{xxxx}-10f_{xxx}^2)\nonumber\\
&+c_1(ff_{txxxxx}-5f_xf_{txxxx}+10f_{xx}f_{txxx}-10f_{xxx}f_{xxt}+5f_{xt}f_{xxxx}-f_tf_{xxxxx})\nonumber\\
&+c_2(ff_{yxxxxx}-5f_xf_{yxxxx}+10f_{xx}f_{yxxx}-10f_{xxx}f_{xxy}+5f_{xy}f_{xxxx}-f_yf_{xxxxx})=0,
\end{align}
and the bilinear form of (\ref{non1}) is
\begin{align}\label{bilinearNON1}
&a_0(ff_{xt}-f_xf_t)+a_1(ff_{yy}-f_y^2)+a_2(ff_{xxxx}-4f_xf_{xxx}+3f_{xx}^2)\nonumber\\
&+b_0(ff_{tyxx}-2f_xf_{xyt}+f_{xx}f_{ty}-f_yf_{txx}+2f_{xt}f_{xy}-f_tf_{yxx})\nonumber\\
&+b_1(ff_{xytt}-2f_tf_{xyt}+f_{tt}f_{xy}-f_yf_{xtt}+2f_{xt}f_{yt}-f_xf_{ytt})\nonumber\\
&+b_2(ff_{xtyy}-2f_yf_{xyt}+f_{yy}f_{xt}-f_tf_{xyy}+2f_{xy}f_{yt}-f_xf_{tyy})=0.
\end{align}

\subsection{Lump solutions with one function}

To obtain lump solutions of (\ref{local1}) and (\ref{non1}) we take $M=1$ in (\ref{Mlump}) i.e.
\begin{equation}\label{1lump}
f=\beta_0+p^2,
\end{equation}
where $p=\beta_1x+\beta_2y+\beta_3t+\beta_4$ for arbitrary constants $\beta_j$, $j=0,1,2,3,4$.

\noindent \textbf{Lump solutions of (\ref{local1}) with one function}

In this part we will obtain lump solutions of the equation (\ref{local1}) with one function. We first insert (\ref{1lump}) into (\ref{bilinearLOC1}) and
get two conditions to be satisfied by $\beta_j$, $j=0,1,2,3,4$ given below.
\begin{align}
&1)\, a_0\beta_3^2+a_2\beta_2^2+a_3\beta_1\beta_3+a_4\beta_2\beta_3+a_5\beta_1\beta_2=0,\label{loc1lumpeq1}\\
&2)\, a_1\beta_1^2\beta_0+6b_0\beta_1^3\beta_2+6b_1\beta_1^3\beta_3+6b_2\beta_1^4=0.\label{loc1lumpeq2}
\end{align}
Let us give a particular example where the solution parameters are satisfying (\ref{loc1lumpeq1}) and (\ref{loc1lumpeq2}).

\noindent \textbf{Example 1.}\, Choose $a_0=1, a_1=-2, a_2=4, a_3=5, a_4=-1, a_5=6, b_0=-4, b_1=2, b_2=3, c_0=5, c_1=2, c_2=-1$ yielding the equation
\begin{align}
&u_{tt}-2u_{xx}+4u_{yy}+5u_{tx}-u_{ty}+6u_{xy}-4(u_{yxx}+3 u_{x} u_{y})_{x} +2(u_{txx}+3 u_{t} u_{x})_{x}
\nonumber\\
&+3(u_{xxx}+3 u_{x}^2)_{x} +5(u_{xxxxx}+15 u_{x} u_{xxx}+ 15 u_{x}^3)_{x} \nonumber \\
&+2(u_{txxxx}+10 u_{x} u_{xxt}+5 u_{t} u_{xxx}+15u_x^2u_t)_{x}
\nonumber \\
&-(u_{yxxxx}+10 u_{x} u_{xxy}+5 u_{y} u_{xx}+15u_x^2u_y)_{x}=0.\label{example1}
\end{align}
Pick also $\beta_0=2, \beta_3=-1, \beta_4=4$ giving $\beta_1=2$ and $\beta_2=1$ obtained from (\ref{loc1lumpeq1}) and (\ref{loc1lumpeq2}). Hence a lump solution of (\ref{example1}) is
\begin{equation}\displaystyle
u(x,y,t)=\frac{8(2x+y-t+4)}{2+(2x+y-t+4)^2}.
\end{equation}
The graphs of the above solution at $t=0$ and $t=10$ with the corresponding contour plots are given in Figure 1.
\begin{center}
\begin{figure}[h!]
    \centering
    \subfigure[]{\includegraphics[width=0.33\textwidth]{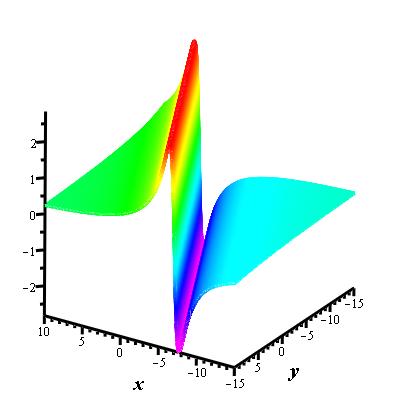}}\hspace{3cm}
    \subfigure[]{\includegraphics[width=0.33\textwidth]{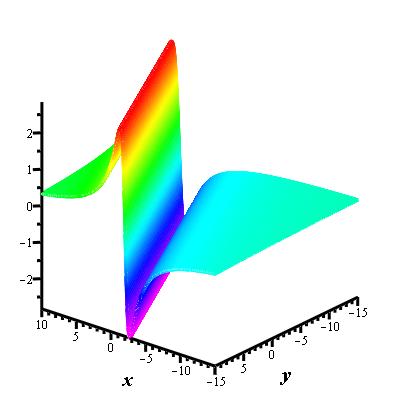}}\hfill
    \subfigure[]{\includegraphics[width=0.33\textwidth]{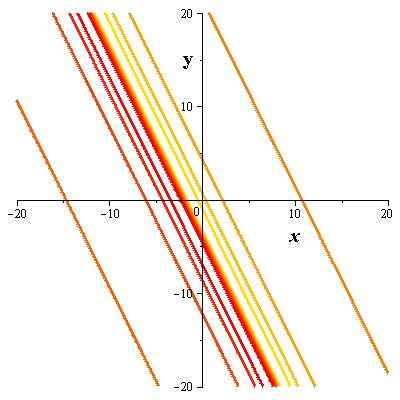}}\hspace{3cm}
    \subfigure[]{\includegraphics[width=0.33\textwidth]{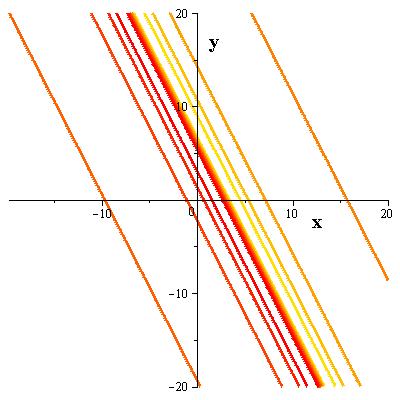}}
    \caption{Lump solutions of (\ref{example1}) at $t=0$ and $t=10$. (a), (c) $t=0$; (b), (d) $t=10$.}
    \end{figure}
\end{center}
%\newpage
\noindent \textbf{Lump solutions of (\ref{non1}) with one function}

Here we consider the equation (\ref{non1}). When we insert (\ref{1lump}) into (\ref{bilinearNON1}) we obtain
two equations to be satisfied by $\beta_j$, $j=0,1,2,3,4$ as
\begin{align}
&1)\, a_0\beta_1\beta_3+a_1\beta_2^2=0,\label{non1lumpeq1}\\
&2)\, a_0\beta_1^2(a_0a_2\beta_1^3-a_1b_0\beta_2^3)+a_1\beta_2^4=0.\label{non1lumpeq2}
\end{align}
We give the following example.

\noindent \textbf{Example 2.}\, Take $a_0=3, a_1=-2, a_2=-18, b_0=-1, b_1=1, b_2=2$ giving the equation
\begin{eqnarray}
&&3u_{xt}-2u_{yy}-18(u_{xxx}+3 u_{x}^2)_{x}-(u_{xxyt}+u_xu_{yt}+2u_yu_{xt}+2u_tu_{xy}+u_{xx}\,D^{-1}\,u_{yt})\nonumber\\
&&+(u_{xytt}+u_yu_{tt}+2u_tu_{yt}+u_{xy}\,D^{-1}\,u_{tt}+2u_{xt}\,D^{-1}\,u_{yt}) \nonumber\\
&&+2(u_{xtyy}+u_tu_{yy}+2u_yu_{yt}+u_{xt}\,D^{-1}\,u_{yy}+2u_{xy}\, D^{-1}\,u_{yt})=0. \label{example2}
\end{eqnarray}
In addition to that choose $\beta_0=2, \beta_1=-3, \beta_4=4$ yielding $\beta_2=9$ and $\beta_3=-18$ from the equations (\ref{non1lumpeq1}) and (\ref{non1lumpeq2}).
Thus a lump solution of (\ref{example2}) is
\begin{equation}\displaystyle
u(x,y,t)=\frac{12(3x-9y+18t-4)}{2+(3x-9y+18t-4)^2}.
\end{equation}
The graphs of the above solution at $t=0$ and $t=2$ with the corresponding contour plots are given in Figure 2.
\begin{center}
\begin{figure}[h!]
    \centering
    \subfigure[]{\includegraphics[width=0.33\textwidth]{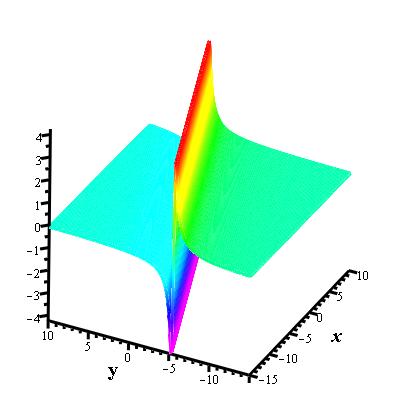}}\hspace{3cm}
    \subfigure[]{\includegraphics[width=0.33\textwidth]{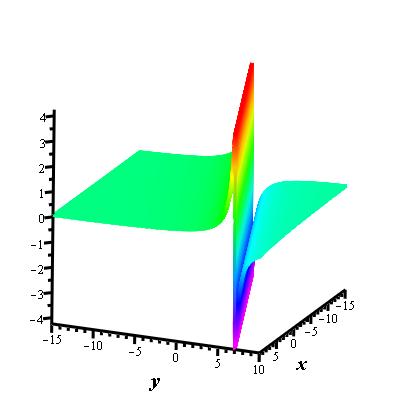}}\hfill
    \subfigure[]{\includegraphics[width=0.33\textwidth]{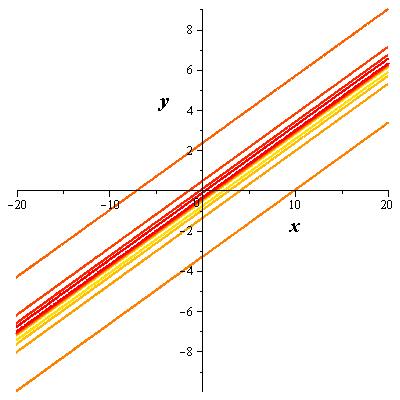}}\hspace{3cm}
    \subfigure[]{\includegraphics[width=0.33\textwidth]{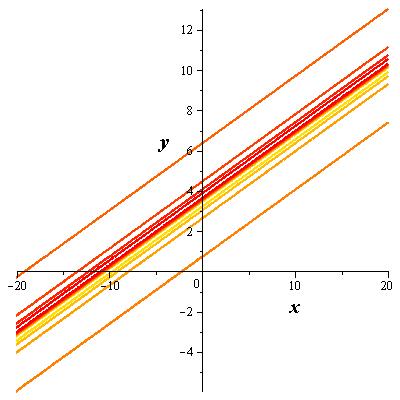}}
    \caption{Lump solutions of (\ref{example2}) at $t=0$ and $t=2$. (a), (c) $t=0$; (b), (d) $t=2$.}
    \end{figure}
\end{center}
\subsection{Lump solutions with two functions}

 To obtain lump solutions of the equations (\ref{local1}) and (\ref{non1}) with two functions we take $M=2$ in (\ref{Mlump}) that is
\begin{equation}\label{2lump}
f=\beta_0+p^2+q^2,
\end{equation}
where $p=\beta_1x+\beta_2y+\beta_3t+\beta_4$ and $q=\beta_5x+\beta_6y+\beta_7t+\beta_8$ for arbitrary constants $\beta_{0}, \beta_{1}, \cdots, \beta_{8}$.
Inserting this ansatz into the bilinear equations of $f$ we obtain some conditions on the constants $\beta_{0}, \beta_{1}, \cdots, \beta_{8}$.

\noindent \textbf{Lump solutions of (\ref{local1}) with two functions}

Inserting (\ref{2lump}) into the bilinear form of the equation (\ref{local1}) yields the following system of equations:
\begin{align}
&1)\, 2\beta_3\beta_7a_0+(\beta_1\beta_6+\beta_2\beta_5)a_5+2\beta_1\beta_5a_1+(\beta_3\beta_5+\beta_1\beta_7)a_3+2\beta_2\beta_6a_2\nonumber\\
&+(\beta_3\beta_6+\beta_2\beta_7)a_4=0,\label{loc2lumpeq1}\\
&2)\,(\beta_2^2-\beta_6^2)a_2+(\beta_3^2-\beta_7^2)a_0+(\beta_1\beta_2-\beta_5\beta_6)a_5+(\beta_1^2-\beta_5^2)a_1+(\beta_1\beta_3-\beta_5\beta_7)a_3\nonumber\\
&+(\beta_2\beta_3-\beta_6\beta_7)a_4=0,\label{loc2lumpeq2}\\
&3)\,6(\beta_1^2+\beta_5^2)^2b_2+\beta_0(\beta_2^2+\beta_6^2)a_2+\beta_0(\beta_3^2+\beta_7^2)a_0+\beta_0(\beta_1^2+\beta_5^2)a_1+\beta_0(\beta_1\beta_2+\beta_5\beta_6)a_5\nonumber\\
&+\beta_0(\beta_1\beta_3+\beta_5\beta_7)a_3+\beta_0(\beta_2\beta_3+\beta_6\beta_7)a_4+6(\beta_1^2+\beta_5^2)(\beta_1\beta_2+\beta_5\beta_6)b_0
\nonumber\\&+6(\beta_1^2+\beta_5^2)(\beta_1\beta_3+\beta_5\beta_7)b_1=0.\label{loc2lumpeq3}
\end{align}
We solve the above system for the constants $\beta_j$, $j=0,1,\cdots,8$ and use them to find lump solutions of (\ref{local1}). Consider the following particular example.

\noindent \textbf{Example 3.}\, Choose $a_0=1, a_1=2, a_2=a_3=1, a_4=-1, a_5=-1, b_0=1, b_1=2$, $b_2=1, c_0=2, c_1=-1, c_2=1$ i.e. we have the equation
\begin{align}
&u_{tt}+2\,u_{xx}+u_{yy}+u_{tx}-u_{ty}-u_{xy}+(u_{yxx}+3 u_{x} u_{y})_{x} +2\,(u_{txx}+3 u_{t} u_{x})_{x}+\,(u_{xxx}+3 u_{x}^2)_{x}\nonumber \\
&+2\,(u_{xxxxx}+15 u_{x} u_{xxx}+ 15 u_{x}^3)_{x}-(u_{txxxx}+10 u_{x} u_{xxt}+5 u_{t} u_{xxx}+15u_x^2u_t)_{x}
\nonumber \\
&+(u_{yxxxx}+10 u_{x} u_{xxy}+5 u_{y} u_{xx}+15u_x^2u_y)_{x}=0.\label{example3}
\end{align}
In addition we pick $\beta_1=2, \beta_2=-1, \beta_4=1, \beta_5=-1, \beta_6=-1, \beta_8=2$ yielding
\begin{equation}\displaystyle
\beta_0=\frac{15\alpha(37+\alpha-\sqrt{1129})}{-4+2\sqrt{1129}},\quad \beta_3=-\frac{3}{2}+\frac{1}{4}\alpha,\quad \beta_7=\frac{10}{\alpha}
\end{equation}
from the equations given in (\ref{loc2lumpeq1})-(\ref{loc2lumpeq3}). Hence we get a lump solution of the equation (\ref{example3})
as{\small
\begin{equation}\displaystyle
u(x,y,t)=\frac{4[5x-y+(-3+\frac{\alpha}{2}-\frac{10}{\alpha})t]}{\frac{15\alpha(37+\alpha-\sqrt{1129})}{-4+2\sqrt{1129}}+(2x-y+(-\frac{3}{2}+\frac{1}{4}\alpha)t+1)^2
+(x+y-\frac{10}{\alpha}t-2)^2}
\end{equation}}
for $\alpha=\sqrt{-54+2\sqrt{1129}}$. The graphs of the above solution at $t=0$ and $t=5$ with the corresponding contour plots are given in Figure 3.
\begin{center}
\begin{figure}[h!]
    \centering
    \subfigure[]{\includegraphics[width=0.32\textwidth]{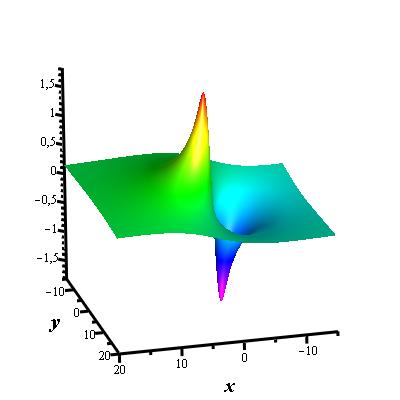}}\hspace{3cm}
    \subfigure[]{\includegraphics[width=0.32\textwidth]{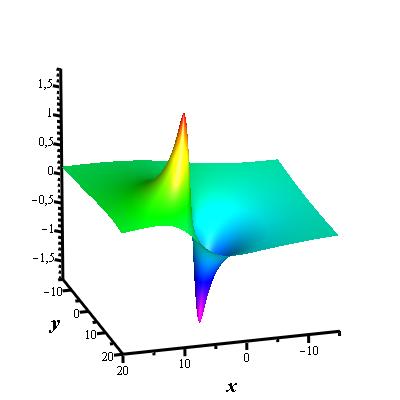}}\hfill
    \subfigure[]{\includegraphics[width=0.32\textwidth]{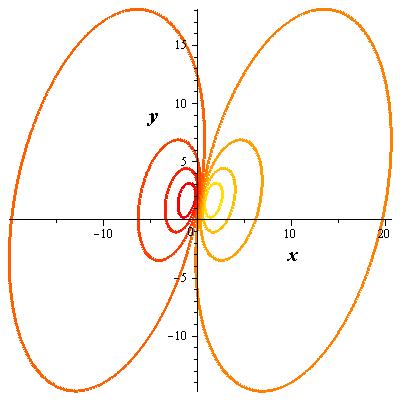}}\hspace{3cm}
    \subfigure[]{\includegraphics[width=0.32\textwidth]{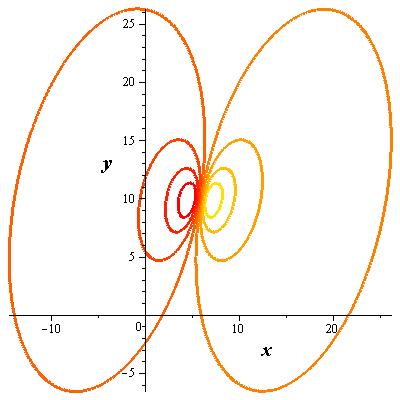}}
    \caption{Lump solutions of (\ref{example3}) at $t=0$ and $t=5$. (a), (c) $t=0$; (b), (d) $t=5$.}
    \end{figure}
\end{center}
\squeezeup
\noindent \textbf{Lump solutions of (\ref{non1}) with two functions}

Inserting the lump solution form (\ref{2lump}) into the bilinear form (\ref{bilinearNON1}) we get the following system of equations to be satisfied by the constants
$\beta_j$, $0\leq j\leq 8$:
\begin{align}
&1)\, a_0(\beta_3\beta_5+\beta_1\beta_7)+2a_1\beta_2\beta_6=0,\label{non2lumpeq1}\\
&2)\, a_1(\beta_2^2-\beta_6^2)+a_0(\beta_1\beta_3-\beta_5\beta_7)=0,\label{non2lumpeq2}\\
&3)\, \beta_0(\beta_2^2+\beta_6^2)a_1+\beta_0(\beta_1\beta_3+\beta_5\beta_7)a_0+6(\beta_1^2+\beta_5^2)^2a_2\nonumber\\
&+2[\beta_2^2(\beta_5\beta_7+3\beta_1\beta_3)+\beta_6^2(3\beta_5\beta_7+\beta_1\beta_3)+2\beta_2\beta_6(\beta_2\beta_7+\beta_3\beta_5)]b_2\nonumber\\
&+2[\beta_2^2(\beta_2\beta_3+3\beta_6\beta_7)+\beta_6^2(3\beta_2\beta_3+\beta_6\beta_7)+2\beta_1\beta_5(\beta_3\beta_6+\beta_2\beta_7)]b_0\nonumber\\
&+2[\beta_2^2(\beta_1\beta_2+3\beta_5\beta_6)+\beta_6^2(3\beta_1\beta_2+\beta_5\beta_6)+2\beta_3\beta_7(\beta_1\beta_6+\beta_2\beta_5)]b_1=0.\label{non2lumpeq3}
\end{align}

Consider the following particular example.

\noindent \textbf{Example 4.}\, Take $a_0=2, a_1=1, a_2=-1, b_0=4, b_1=5, b_2=-2$ yielding the equation
\begin{eqnarray}
&&2u_{xt}+u_{yy}-(u_{xxx}+3 u_{x}^2)_{x}+4(u_{xxyt}+u_xu_{yt}+2u_yu_{xt}+2u_tu_{xy}+u_{xx}\,D^{-1}\,u_{yt})\nonumber\\
&&+5(u_{xytt}+u_yu_{tt}+2u_tu_{yt}+u_{xy}\,D^{-1}\,u_{tt}+2u_{xt}\,D^{-1}\,u_{yt}) \nonumber\\
&&-2(u_{xtyy}+u_tu_{yy}+2u_yu_{yt}+u_{xt}\,D^{-1}\,u_{yy}+2u_{xy}\, D^{-1}\,u_{yt})=0.\label{example4}
\end{eqnarray}
We choose also $\beta_1=2, \beta_2=-1, \beta_4=2, \beta_5=-1, \beta_6=-1, \beta_8=2$ giving $\beta_0=44, \beta_3=\frac{1}{5}, \beta_7=-\frac{2}{5}$ from the equations (\ref{non2lumpeq1})-(\ref{non2lumpeq3}). Hence we obtain a lump solution of the equation (\ref{example4})
as
\begin{equation}\displaystyle
u=\frac{4(5x-y+\frac{4}{5}t+2)}{44+(2x-y+\frac{1}{5}t+2)^2+(x+y+\frac{2}{5}t-2)^2}.
\end{equation}
The graphs of the above solution at $t=0$ and $t=60$ with the corresponding contour plots are given in Figure 4.
\begin{center}
\begin{figure}[h!]
    \centering
    \subfigure[]{\includegraphics[width=0.33\textwidth]{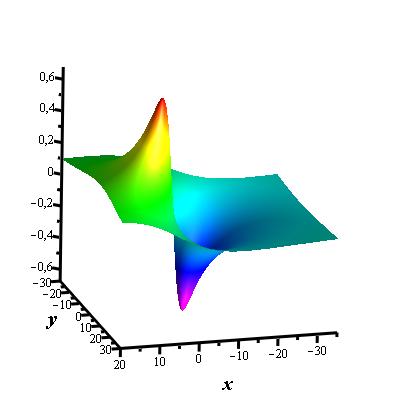}}\hspace{3cm}
    \subfigure[]{\includegraphics[width=0.33\textwidth]{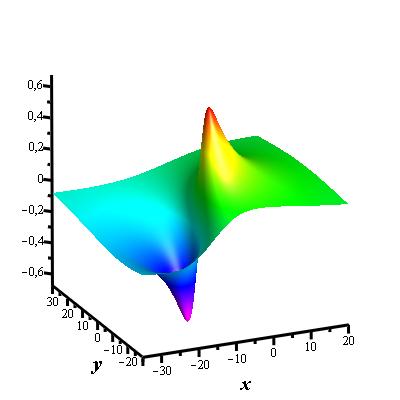}}\hfill
    \subfigure[]{\includegraphics[width=0.33\textwidth]{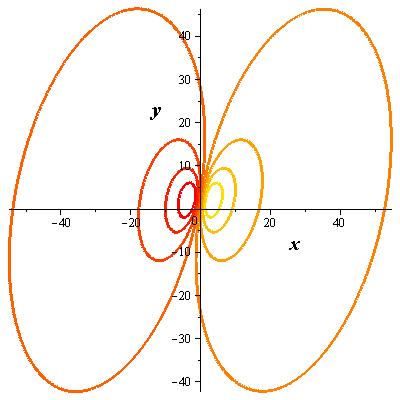}}\hspace{3cm}
    \subfigure[]{\includegraphics[width=0.33\textwidth]{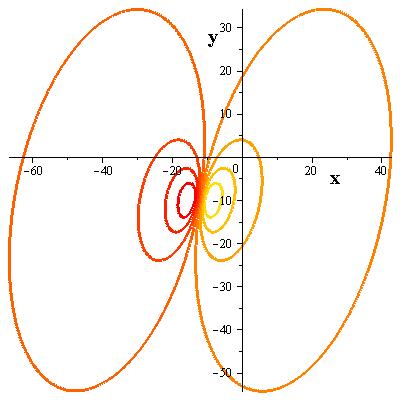}}
    \caption{Lump solutions of (\ref{example4}) at $t=0$ and $t=60$. (a), (c) $t=0$; (b), (d) $t=60$.}
    \end{figure}
\end{center}

\subsection{Lump solutions with three functions}

We can consider a more general form for the function $f$ by taking $M=3$ in (\ref{Mlump}) \cite{Ma-0}, \cite{Ma-7}, \cite{BaiDeng}
\begin{equation}\label{3lump}
f=\beta_0+p^2+q^2+r^2,
\end{equation}
with $p=\beta_1 x+\beta_2 y+\beta_3 t+\beta_4$, $q=\beta_5 x+\beta_6 y+\beta_7 t+\beta_8$, and $r=\beta_9 x+\beta_{10} y+\beta_{11} t+\beta_{12}$  where $\beta_{0}, \beta_{1}, \cdots, \beta_{12}$ are arbitrary constants.
Similar to the previous section inserting the above ansatz into the bilinear equations of (\ref{bilinearLOC1}) and (\ref{bilinearNON1}), we obtain systems of equations for the constants $\beta_{0}, \beta_{1}, \cdots, \beta_{12}$ given in Appendix B.

For the equation (\ref{local1}) the relations given in Appendix B result in
\begin{align}
&p=\beta_3(\eta x+\xi y+t)+\beta_4,\\
&q=\beta_7(\eta x+\xi y+t)+\beta_8,\\
&r=\beta_{11}(\eta x+\xi y+t)+\beta_{12},
\end{align}
where $\xi=\frac{\beta_{10}}{\beta_{11}}$ and $\eta$ is given by (\ref{3luloceta}). In this case we have a lump solution to (\ref{local1}) similar to
a lump solution with one function.

For the equation (\ref{non1}) the relations given in Appendix B yields
\begin{align}
&p=\beta_3(-\lambda\mu^2x+\mu y+t)+\beta_4,\\
&q=\beta_7(-\lambda\mu^2x+\mu y+t)+\beta_8,\\
&r=\beta_{11}(-\lambda\mu^2x+\mu y+t)+\beta_{12},
\end{align}
which gives a lump solution to (\ref{non1}) similar to a lump solution with one function.

We now give particular examples for lump solutions of (\ref{local1}) and (\ref{non1}) with three functions.

\noindent \textbf{Example 5.} Take the coefficients of the equation (\ref{local1}) as $a_0=1, a_1=-2, a_2=1, a_3=2, a_4=-1, a_5=1, b_0=3, b_1=2, b_2=-1, c_0=1, c_1=-2, c_2=1$ yielding
\begin{eqnarray}
&&u_{tt}-2\,u_{xx}+u_{yy}+2u_{tx}-u_{ty}+u_{xy}+3\,(u_{yxx}+3 u_{x} u_{y})_{x} \nonumber\\
&&+2\,(u_{txx}+3 u_{t} u_{x})_{x}-(u_{xxx}+3 u_{x}^2)_{x} +(u_{xxxxx}+15 u_{x} u_{xxx}+ 15 u_{x}^3)_{x} \nonumber \\
&&-2\,(u_{txxxx}+10 u_{x} u_{xxt}+5 u_{t} u_{xxx}+15u_x^2u_t)_{x}
\nonumber \\
&&+\,(u_{yxxxx}+10 u_{x} u_{xxy}+5 u_{y} u_{xx}+15u_x^2u_y)_{x}=0.\label{example5}
\end{eqnarray}
In addition to that if we pick $\beta_3=2, \beta_5=1, \beta_{10}=-3$ we obtain $\beta_1=-2, \beta_2=-2, \beta_6=1, \beta_7=-1, \beta_9=-3,$ and $\beta_{11}=3$.
Choose also $\beta_0=2$, $\beta_4=-1$, $\beta_8=4$, and $\beta_{12}=1$. Hence a lump solution of the equation (\ref{example5}) is
{\small\begin{equation}\displaystyle
u(x,y,t)=\frac{4[14(x+y-t)+3]}{2+(2[x+y-t]+1)^2+([x+y-t]+4)^2+(3[x+y-t]-1)^2}.
\end{equation}}
The graphs of the above solution at $t=0$ and $t=20$ with the corresponding contour plots are given in Figure 5.
\begin{center}
\begin{figure}[h!]
    \centering
    \subfigure[]{\includegraphics[width=0.32\textwidth]{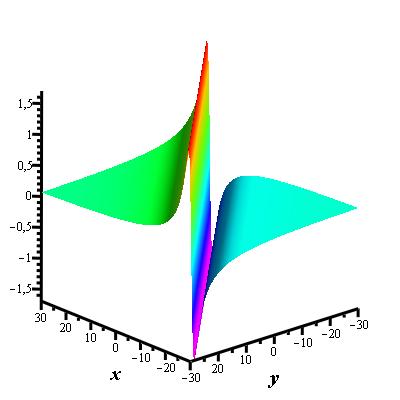}}\hspace{3cm}
    \subfigure[]{\includegraphics[width=0.32\textwidth]{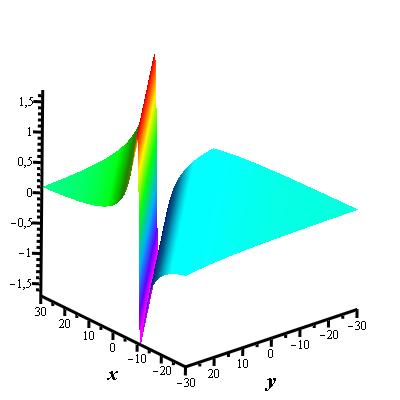}}\hfill
    \subfigure[]{\includegraphics[width=0.32\textwidth]{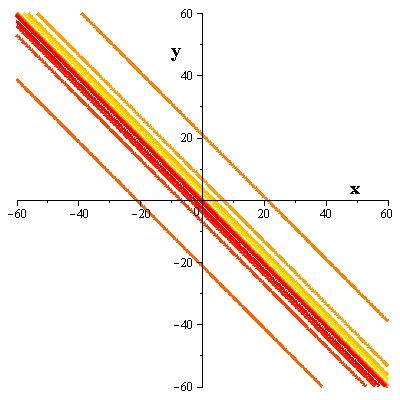}}\hspace{3cm}
    \subfigure[]{\includegraphics[width=0.32\textwidth]{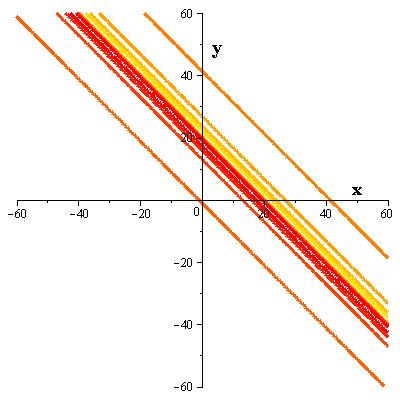}}
    \caption{Lump solutions of (\ref{example5}) at $t=0$ and $t=20$. (a), (c) $t=0$; (b), (d) $t=20$.}
    \end{figure}
\end{center}
\newpage
\noindent \textbf{Example 6.} Choose the coefficients of the equation (\ref{non1}) as $a_0=1, a_1=3, a_2=6, b_0=16, b_1=2, b_2=3$ yielding
\begin{eqnarray}
&&u_{xt}+3u_{yy}+6(u_{xxx}+3 u_{x}^2)_{x}+16(u_{xxyt}+u_xu_{yt}+2u_yu_{xt}+2u_tu_{xy}+u_{xx}\,D^{-1}\,u_{yt})\nonumber\\
&&+2(u_{xytt}+u_yu_{tt}+2u_tu_{yt}+u_{xy}\,D^{-1}\,u_{tt}+2u_{xt}\,D^{-1}\,u_{yt}) \nonumber\\
&&+3(u_{xtyy}+u_tu_{yy}+2u_yu_{yt}+u_{xt}\,D^{-1}\,u_{yy}+2u_{xy}\, D^{-1}\,u_{yt})=0. \label{example6}
\end{eqnarray}
We take $\beta_3=1, \beta_7=-5, \beta_{10}=2, \beta_{11}=-3$ yielding $\beta_1=-\frac{4}{3}, \beta_2=-\frac{2}{3}, \beta_5=\frac{20}{3}, \beta_6=\frac{10}{3}, \beta_9=4$ from
the relations given in Appendix B.
Choose also $\beta_0=1, \beta_4=2, \beta_8=-1$, and $\beta_{12}=7$. Hence we obtain a lump solution of the equation (\ref{example6}) as
\begin{equation}\displaystyle
u(x,y,t)=\frac{16(140x+70y-105t+42)}{9(1+(\frac{4}{3}x+\frac{2}{3}y-t-2)^2+(\frac{20}{3}x+\frac{10}{3}y-5t-1)^2+(4x+2y-3t+7)^2)}.
\end{equation}
The graphs of the above solution at $t=0$ and $t=20$ with the corresponding contour plots are given in Figure 6.
\begin{center}
\begin{figure}[h!]
    \centering
    \subfigure[]{\includegraphics[width=0.33\textwidth]{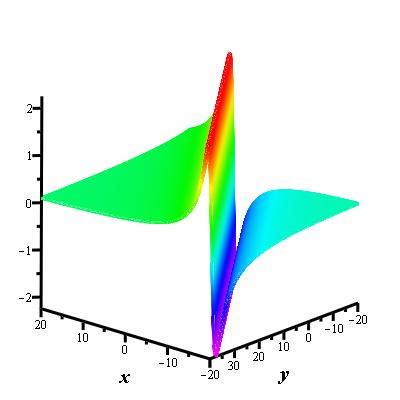}}\hspace{3cm}
    \subfigure[]{\includegraphics[width=0.33\textwidth]{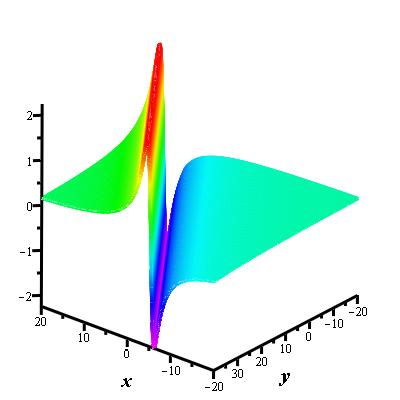}}\hfill
    \subfigure[]{\includegraphics[width=0.33\textwidth]{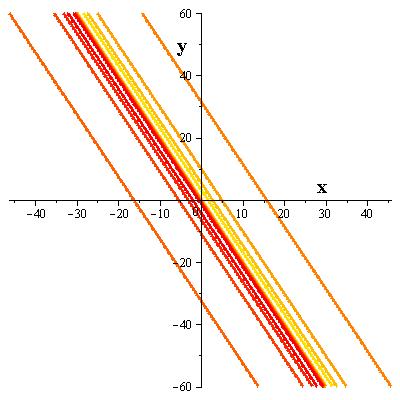}}\hspace{3cm}
    \subfigure[]{\includegraphics[width=0.33\textwidth]{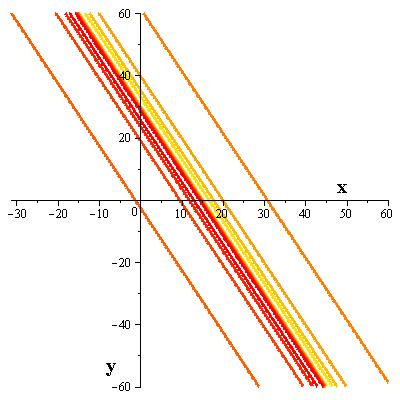}}
    \caption{Lump solutions of (\ref{example6}) at $t=0$ and $t=20$. (a), (c) $t=0$; (b), (d) $t=20$.}
    \end{figure}
\end{center}

All the lump solutions of  (\ref{local1}) and (\ref{non1}) are regular everywhere and asymptotically decaying to zero.

\section{Hybrid solutions}

\noindent \textbf{Case 1.} Consider the following ansatz
\begin{equation}
f=\beta_0+p^2+q,
\end{equation}
where $p=\beta_1x+\beta_2y+\beta_3t+\beta_4$, $q=e^{\beta_5x+\beta_6y+\beta_7t+\beta_8}$ with arbitrary constants $\beta_{0}, \beta_{1}, \cdots, \beta_8$.
Inserting the above ansatz into the bilinear equations (\ref{bilinearLOC1}) and (\ref{bilinearNON1}), we obtain systems of equations for the constants $\beta_{0}, \beta_{1}, \cdots, \beta_{8}$ given in Appendix C.

For the equation (\ref{local1}) due to the condition (\ref{case1loc1}) we have two possibilities:
\begin{equation}
i)\, \beta_1=0,\quad \quad ii)\, \beta_1\neq 0,\,\, \beta_3=-\frac{(b_2\beta_1+b_0\beta_2)}{b_1}.
\end{equation}
Let $\beta_1=0$. Consider the following example for this case.

\noindent \textbf{Example 7.} Take the coefficients of the equation (\ref{local1}) as $a_0=1, a_1=1, a_2=-3, a_3=1, a_4=2, a_5=b_0=b_1=1, b_2=-1, c_0=c_1=1, c_2=-1$ giving
\begin{eqnarray}
&&u_{tt}+u_{xx}-3u_{yy}+u_{tx}+2u_{ty}+u_{xy}+(u_{yxx}+3 u_{x} u_{y})_{x} \nonumber\\
&&+(u_{txx}+3 u_{t} u_{x})_{x}-(u_{xxx}+3 u_{x}^2)_{x} +(u_{xxxxx}+15 u_{x} u_{xxx}+ 15 u_{x}^3)_{x} \nonumber\\
&&+(u_{txxxx}+10 u_{x} u_{xxt}+5 u_{t} u_{xxx}+15u_x^2u_t)_{x}\nonumber\\
&&-(u_{yxxxx}+10 u_{x} u_{xxy}+5 u_{y} u_{xx}+15u_x^2u_y)_{x}=0.\label{example7}
\end{eqnarray}
Pick also $\beta_0=1, \beta_3=1, \beta_4=-2, \beta_7=3$, $\beta_8=2$ yielding $\beta_2=-\frac{1}{3}, \beta_5=-\frac{\sqrt{2}}{2}, \beta_6=\frac{1}{6}\sqrt{2}-1$.
Hence a lump solution of the equation (\ref{example7}) is
\begin{equation}\displaystyle
u(x,y,t)=\frac{\sqrt{2}e^{\frac{-\sqrt{2}}{2}x+(\frac{1}{6}\sqrt{2}-1)y+3t+2} }{1+(-2-\frac{1}{3}y+t)^2+e^{\frac{-\sqrt{2}}{2}x+(\frac{1}{6}\sqrt{2}-1)y+3t+2}}.
\end{equation}
The graphs of the above solution at $t=0$ and $t=5$ with the corresponding contour plots are given in Figure 7.
\begin{center}
\begin{figure}[h!]
    \centering
    \subfigure[]{\includegraphics[width=0.33\textwidth]{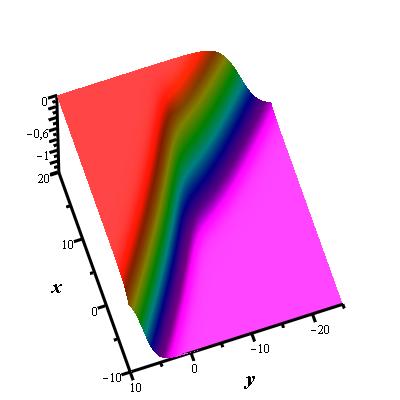}}\hspace{3cm}
    \subfigure[]{\includegraphics[width=0.33\textwidth]{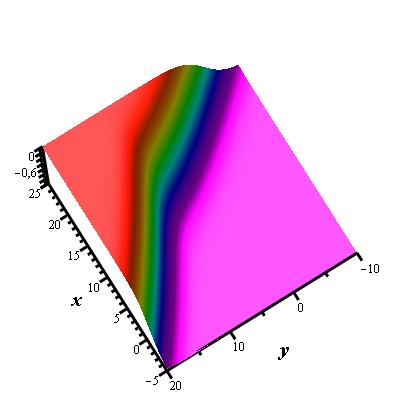}}\hfill
    \subfigure[]{\includegraphics[width=0.33\textwidth]{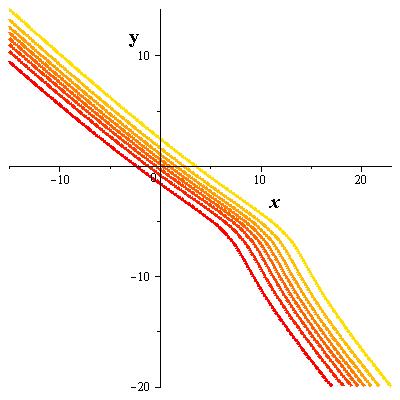}}\hspace{3cm}
    \subfigure[]{\includegraphics[width=0.33\textwidth]{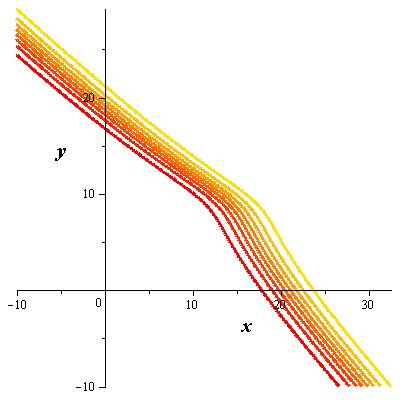}}
    \caption{Lump solutions of (\ref{example7}) at $t=0$ and $t=5$. (a), (c) $t=0$; (b), (d) $t=5$.}
    \end{figure}
\end{center}
Now we will consider the case when $\beta_1\neq 0, \beta_3=-\frac{(b_2\beta_1+b_0\beta_2)}{b_1}$ for the equation (\ref{local1}) and give
the following example.

\noindent \textbf{Example 8.} Choose the coefficients of the equation (\ref{local1}) as $a_0=a_1=1, a_2=-1, a_3=a_4=a_5=b_0=b_1=b_2=c_0=c_1=c_2=1$. Hence we have the equation
\begin{eqnarray}
&&u_{tt}+u_{xx}-u_{yy}+u_{tx}+u_{ty}+u_{xy}+(u_{yxx}+3 u_{x} u_{y})_{x} \nonumber\\
&&+(u_{txx}+3 u_{t} u_{x})_{x}-(u_{xxx}+3 u_{x}^2)_{x} +(u_{xxxxx}+15 u_{x} u_{xxx}+ 15 u_{x}^3)_{x} \nonumber\\
&&+(u_{txxxx}+10 u_{x} u_{xxt}+5 u_{t} u_{xxx}+15u_x^2u_t)_{x}\nonumber\\
&&+(u_{yxxxx}+10 u_{x} u_{xxy}+5 u_{y} u_{xx}+15u_x^2u_y)_{x}=0.\label{example8}
\end{eqnarray}
We take also $\beta_0=1, \beta_3=1, \beta_4=-2, \beta_6=2$, $\beta_8=4$ yielding $\beta_1=-\frac{3}{2}+\frac{1}{2}\sqrt{5}$, $\beta_2=\frac{1}{2}-\frac{1}{2}\sqrt{5}$, $\beta_5=-1+\sqrt{5}$, and $\beta_7=-1-\sqrt{5}$.
Therefore a lump solution of the equation (\ref{example8}) is $u(x,y,t)=\frac{U(x,y,t)}{V(x,y,t)}$, where
{\small\begin{align}
U(x,y,t)&=2[2(-\frac{3}{2}+\frac{1}{2}\sqrt{5})((-\frac{3}{2}+\frac{1}{2}\sqrt{5})x+(\frac{1}{2}-\frac{1}{2}\sqrt{5})y+t-2)\nonumber\\
&+(-1+\sqrt{5})e^{(-1+\sqrt{5})x+2y-(1+\sqrt{5})t+4} ],\\
V(x,y,t)&=1+((-\frac{3}{2}+\frac{1}{2}\sqrt{5})x+(\frac{1}{2}-\frac{1}{2}\sqrt{5})y+t-2)^2+e^{(-1+\sqrt{5})x+2y-(1+\sqrt{5})t+4}.
\end{align}}
The graphs of the above solution at $t=0$ and $t=5$ with the corresponding contour plots are given in Figure 8.
\squeezeup
\begin{center}
\begin{figure}[h!]
    \centering
    \subfigure[]{\includegraphics[width=0.32\textwidth]{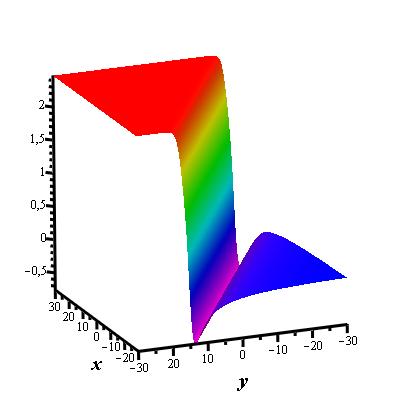}}\hspace{3cm}
    \subfigure[]{\includegraphics[width=0.32\textwidth]{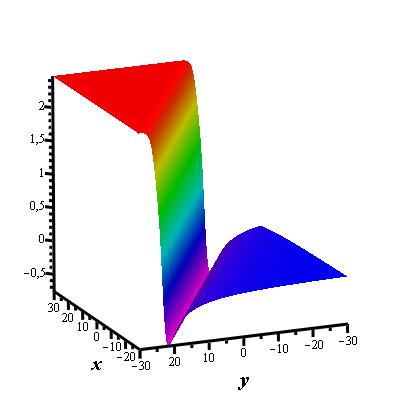}}\hfill
    \subfigure[]{\includegraphics[width=0.32\textwidth]{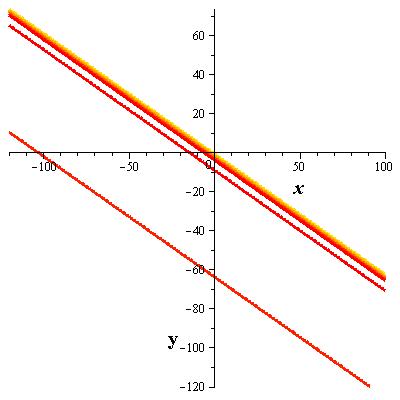}}\hspace{3cm}
    \subfigure[]{\includegraphics[width=0.32\textwidth]{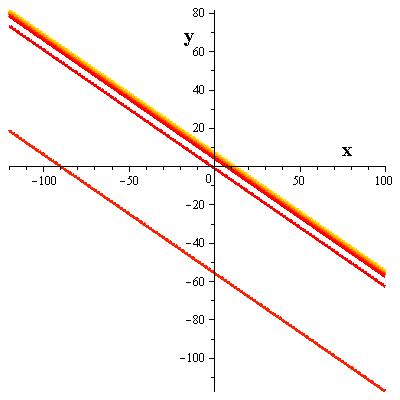}}
    \caption{Lump solutions of (\ref{example8}) at $t=0$ and $t=5$. (a), (c) $t=0$; (b), (d) $t=6$.}
    \end{figure}
\end{center}
For the equation (\ref{non1}) by following the constraints for the parameters given in Appendix C, we give the following example.

\noindent \textbf{Example 9.} Let $a_0=-2, a_1=a_2=1, b_0=-1, b_1=4, b_2=-3$ in (\ref{non1}). We have the equation
\begin{eqnarray}
&&-2u_{xt}+u_{yy}+(u_{xxx}+3 u_{x}^2)_{x}-(u_{xxyt}+u_xu_{yt}+2u_yu_{xt}+2u_tu_{xy}+u_{xx}\,D^{-1}\,u_{yt})\nonumber\\
&&+4(u_{xytt}+u_yu_{tt}+2u_tu_{yt}+u_{xy}\,D^{-1}\,u_{tt}+2u_{xt}\,D^{-1}\,u_{yt}) \nonumber\\
&&-3(u_{xtyy}+u_tu_{yy}+2u_yu_{yt}+u_{xt}\,D^{-1}\,u_{yy}+2u_{xy}\, D^{-1}\,u_{yt})=0, \label{example9}
\end{eqnarray}
Take also $\beta_0=2, \beta_1=1, \beta_4=-1, \beta_5=3, \beta_8=1$ yielding $\beta_2=1, \beta_3=\frac{1}{2}, \beta_6=3, \beta_7=\frac{3}{2}$.
Hence a lump solution of the equation (\ref{example9}) is
\begin{equation}\displaystyle
u(x,y,t)=\frac{2[2(x+y+\frac{1}{2}t-1)+3e^{3x+3y+\frac{3}{2}t+1}]}{2+(x+y+\frac{1}{2}t-1)^2+e^{3x+3y+\frac{3}{2}t+1}}.
\end{equation}
The graphs of the above solution at $t=0$ and $t=10$ with the corresponding contour plots are given in Figure 9.
\begin{center}
\begin{figure}[h!]
    \centering
    \subfigure[]{\includegraphics[width=0.32\textwidth]{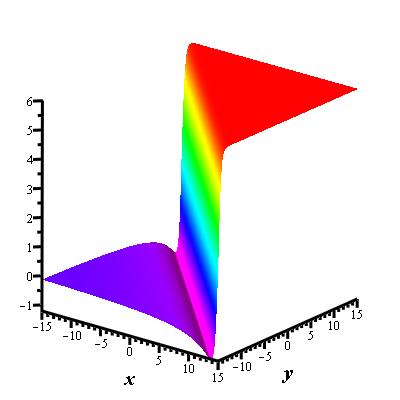}}\hspace{3cm}
    \subfigure[]{\includegraphics[width=0.32\textwidth]{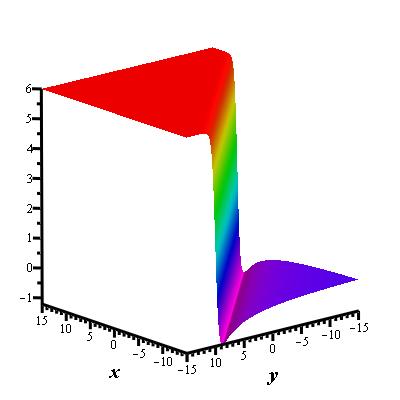}}\hfill
    \subfigure[]{\includegraphics[width=0.32\textwidth]{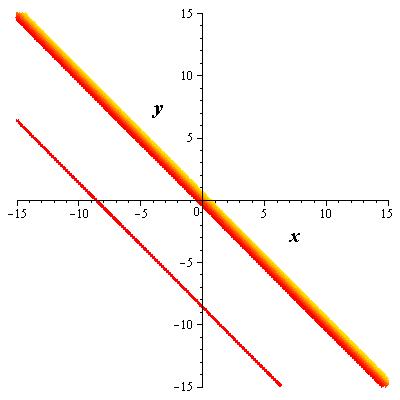}}\hspace{3cm}
    \subfigure[]{\includegraphics[width=0.31\textwidth]{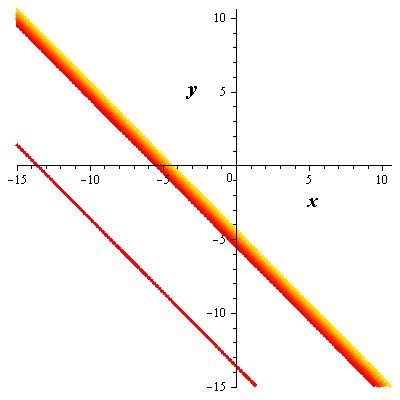}}
    \caption{Lump solutions of (\ref{example9}) at $t=0$ and $t=10$. (a), (c) $t=0$; (b), (d) $t=10$.}
    \end{figure}
\end{center}

\noindent \textbf{Case 2.} Here we consider
\begin{equation}
f=\beta_0+p^2+q+Ap^2q,
\end{equation}
where $p=\beta_1x+\beta_2y+\beta_3t+\beta_4$, $q=e^{\beta_5x+\beta_6y+\beta_7t+\beta_8}$ with arbitrary constants $\beta_{0}, \beta_{1}, \cdots, \beta_8, A$.
Inserting the above ansatz into the bilinear equations of (\ref{bilinearLOC1}) and (\ref{bilinearNON1}), we obtain systems of equations for the constants $\beta_{0}, \beta_{1}, \cdots, \beta_{8}, A$ given in Appendix C. If $A=0$ then this case turns to be the Case 1. Hence here we only consider when $A\neq 0$.

For the equation (\ref{local1}) here also we have two possibilities:
\begin{equation}
i)\, \beta_1=0,\quad \quad ii)\, \beta_1\neq 0,\,\, \beta_3=-\frac{(b_2\beta_1+b_0\beta_2)}{b_1}.
\end{equation}

Let us first give an example corresponding to $\beta_1=0$.
\newpage

\noindent \textbf{Example 10.} Take the coefficients of the equation (\ref{local1}) as $a_0=a_1=1, a_2=-3$, $a_3=1$, $a_4=2$, $a_5=b_0=b_1=1$, $b_2=-1$, $c_0=c_1=1, c_2=-1$. Hence we have the equation
\begin{eqnarray}
&&u_{tt}+u_{xx}-3u_{yy}+u_{tx}+2u_{ty}+u_{xy}+(u_{yxx}+3 u_{x} u_{y})_{x} \nonumber\\
&&+(u_{txx}+3 u_{t} u_{x})_{x}-(u_{xxx}+3 u_{x}^2)_{x} +(u_{xxxxx}+15 u_{x} u_{xxx}+ 15 u_{x}^3)_{x} \nonumber\\
&&+(u_{txxxx}+10 u_{x} u_{xxt}+5 u_{t} u_{xxx}+15u_x^2u_t)_{x}\nonumber\\
&&-(u_{yxxxx}+10 u_{x} u_{xxy}+5 u_{y} u_{xx}+15u_x^2u_y)_{x}=0.\label{example10}
\end{eqnarray}
Choose also $\beta_0=1, \beta_2=3, \beta_4=-2, \beta_6=1, \beta_8=2, A=4$ yielding $\beta_3=3, \beta_5=\frac{\sqrt{2}}{2}, \beta_7=1-\frac{3}{8}\sqrt{2}$.
Thus a lump solution of the equation (\ref{example10}) is
\begin{equation}\displaystyle
u(x,y,t)=\frac{2[\frac{\sqrt{2}}{2}e^{\frac{\sqrt{2}}{2}x+y+(1-\frac{3}{8}\sqrt{2})t+2}+2(-2+3y+3t)^2\sqrt{2}e^{\frac{\sqrt{2}}{2}x+y+(1-\frac{3}{8}\sqrt{2})t+2}  ]}{1+(-2+3y+3t)^2+e^{\frac{\sqrt{2}}{2}x+y+(1-\frac{3}{8}\sqrt{2})t+2}+4(-2+3y+3t)^2e^{\frac{\sqrt{2}}{2}x+y+(1-\frac{3}{8}\sqrt{2})t+2}}.
\end{equation}
The graphs of the above solution at $t=0$ and $t=5$ with the corresponding contour plots are given in Figure 10.
\begin{center}
\begin{figure}[h!]
    \centering
    \subfigure[]{\includegraphics[width=0.33\textwidth]{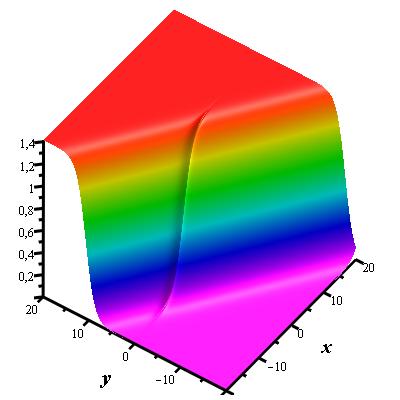}}\hspace{3cm}
    \subfigure[]{\includegraphics[width=0.33\textwidth]{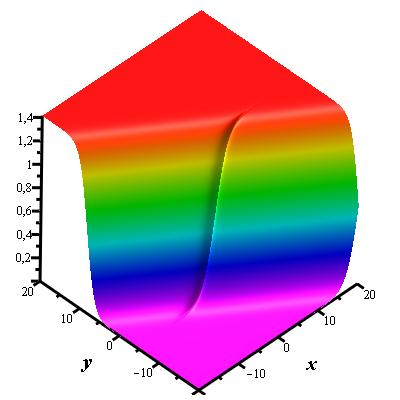}}\hfill
    \subfigure[]{\includegraphics[width=0.33\textwidth]{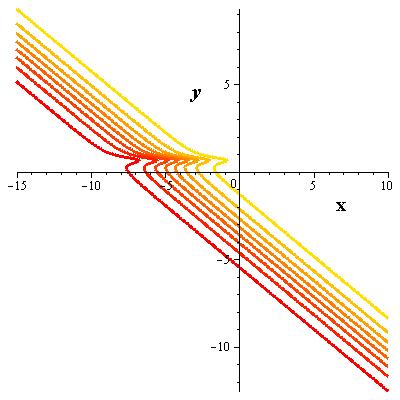}}\hspace{3cm}
    \subfigure[]{\includegraphics[width=0.33\textwidth]{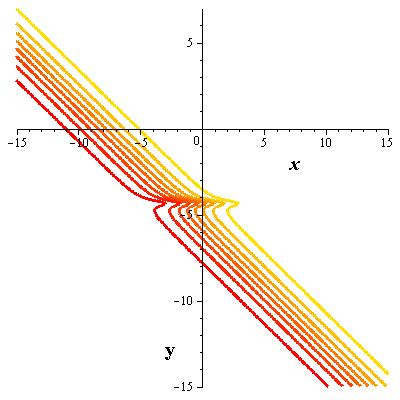}}
    \caption{Lump solutions of (\ref{example10}) at $t=0$ and $t=5$. (a), (c) $t=0$; (b), (d) $t=5$.}
    \end{figure}
\end{center}
Now consider the case when $\beta_1\neq 0, \beta_3=-\frac{(b_2\beta_1+b_0\beta_2)}{b_1}$. By following the constraints given in Appendix C, we give the following example.

\noindent \textbf{Example 11.} Take the coefficients of the equation (\ref{local1}) as $a_0=a_1=1, a_2=-2, a_3=a_4=a_5=b_0=b_1=b_2=c_0=c_1=1, c_2=-2$. Hence we have the equation
\begin{eqnarray}
&&u_{tt}+u_{xx}-2u_{yy}+u_{tx}+u_{ty}+u_{xy}+(u_{yxx}+3 u_{x} u_{y})_{x} \nonumber\\
&&+(u_{txx}+3 u_{t} u_{x})_{x}+(u_{xxx}+3 u_{x}^2)_{x} +(u_{xxxxx}+15 u_{x} u_{xxx}+ 15 u_{x}^3)_{x} \nonumber\\
&&+(u_{txxxx}+10 u_{x} u_{xxt}+5 u_{t} u_{xxx}+15u_x^2u_t)_{x}\nonumber\\
&&-2(u_{yxxxx}+10 u_{x} u_{xxy}+5 u_{y} u_{xx}+15u_x^2u_y)_{x}=0.\label{example11}
\end{eqnarray}
Pick also $\beta_0=\frac{1}{2}, \beta_1=-1, \beta_4=-2, \beta_8=2, A=2$ yielding $\beta_2=\frac{1}{2}, \beta_3=\frac{1}{2}, \beta_5=\frac{3}{59}\sqrt{295}, \beta_6=\frac{1029}{6962}\sqrt{295}$, and $\beta_7=\frac{321}{3481}\sqrt{295}$.
Hence a lump solution of the equation (\ref{example11}) is
\begin{equation}
u(x,y,t)=\frac{U(x,y,t)}{V(x,y,t)},
\end{equation}
where
\begin{align}
U(x,y,t)&=2[-2(-x+\frac{1}{2}y+\frac{1}{2}t-2)+\frac{3}{59}\sqrt{295}e^{\frac{3}{59}\sqrt{295}x+\frac{1029}{6962}\sqrt{295}y+\frac{321}{3481}\sqrt{295}t+2}\nonumber\\
&-4(-x+\frac{1}{2}y+\frac{1}{2}t-2)e^{\frac{3}{59}\sqrt{295}x+\frac{1029}{6962}\sqrt{295}y+\frac{321}{3481}\sqrt{295}t+2}
\nonumber\\&+\frac{6\sqrt{295}}{59}(-x+\frac{1}{2}y+\frac{1}{2}t-2)^2e^{\frac{3}{59}\sqrt{295}x+\frac{1029}{6962}\sqrt{295}y+\frac{321}{3481}\sqrt{295}t+2}],\\
V(x,y,t)&=\frac{1}{2}+(-x+\frac{1}{2}y+\frac{1}{2}t-2)^2+e^{\frac{3}{59}\sqrt{295}x+\frac{1029}{6962}\sqrt{295}y+\frac{321}{3481}\sqrt{295}t+2}\nonumber
\\&+2(-x+\frac{1}{2}y+\frac{1}{2}t-2)^2e^{\frac{3}{59}\sqrt{295}x+\frac{1029}{6962}\sqrt{295}y+\frac{321}{3481}\sqrt{295}t+2}
\end{align}
The graphs of the above solution at $t=0$ and $t=10$ with the corresponding contour plots are given in Figure 11.
\begin{center}
\begin{figure}[h!]
    \centering
    \subfigure[]{\includegraphics[width=0.33\textwidth]{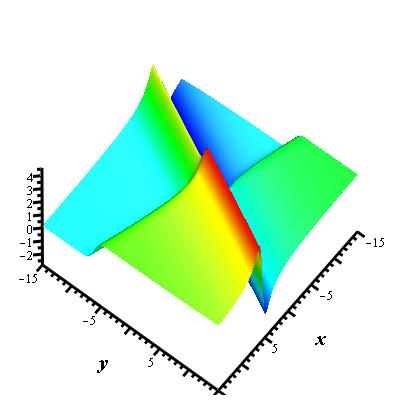}}\hspace{3cm}
    \subfigure[]{\includegraphics[width=0.33\textwidth]{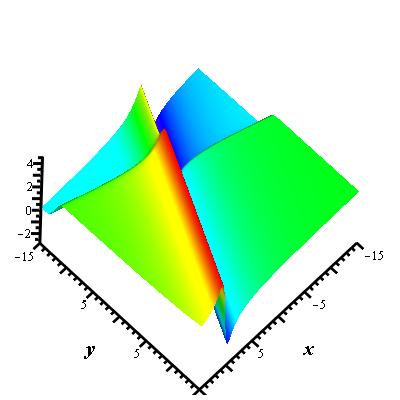}}\hfill
    \subfigure[]{\includegraphics[width=0.33\textwidth]{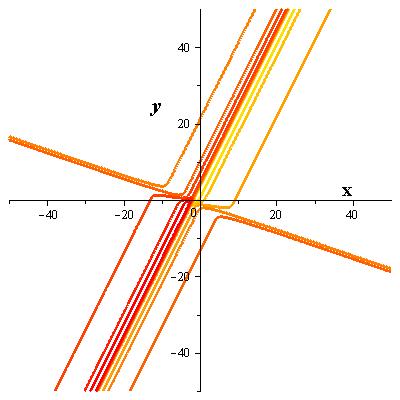}}\hspace{3cm}
    \subfigure[]{\includegraphics[width=0.33\textwidth]{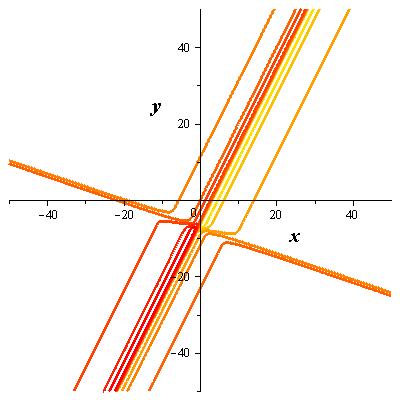}}
    \caption{Lump solutions of (\ref{example11}) at $t=0$ and $t=10$. (a), (c) $t=0$; (b), (d) $t=10$.}
    \end{figure}
\end{center}
\newpage
\noindent For the equation (\ref{non1}) we present the following example.

\noindent \textbf{Example 12.} Note that the parameters satisfying the conditions of Case 1 for (\ref{non1}) given in Appendix C also satisfy
the conditions of Case 2. Therefore we can consider the same equation (\ref{example9}) with same $\beta_j$, $j=0,\cdots, 8$. In addition to that let us pick $A=10$.
Hence we find another lump solution $u(x,y,t)=\frac{U(x,y,t)}{V(x,y,t)}$ of (\ref{example9}) where
{\small
\begin{align}\displaystyle
U(x,y,t)&=2[2(x+y+\frac{t}{2}-1)+3e^{3x+3y+\frac{3}{2}t+1}+20(x+y+\frac{t}{2}-1)e^{3x+3y+\frac{3}{2}t+1}\nonumber\\
&+30(x+y+\frac{t}{2}-1)^2e^{3x+3y+\frac{3}{2}t+1}],\\
V(x,y,t)&=2+(x+y+\frac{t}{2}-1)^2+e^{3x+3y+\frac{3}{2}t+1}+10(x+y+\frac{t}{2}-1)^2e^{3x+3y+\frac{3}{2}t+1}.
\end{align}}
The graphs of the above solution at $t=0$ and $t=10$ with the corresponding contour plots are given in Figure 12.
\begin{center}
\begin{figure}[h!]
    \centering
    \subfigure[]{\includegraphics[width=0.32\textwidth]{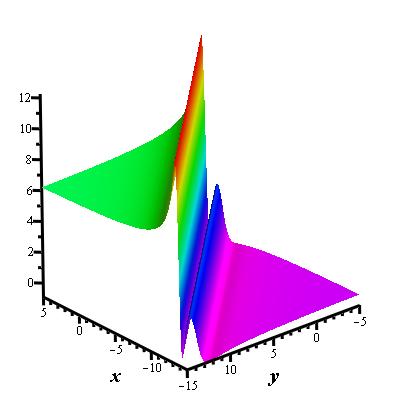}}\hspace{3cm}
    \subfigure[]{\includegraphics[width=0.32\textwidth]{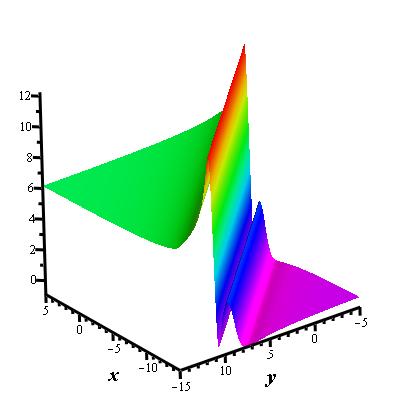}}\hfill
    \subfigure[]{\includegraphics[width=0.32\textwidth]{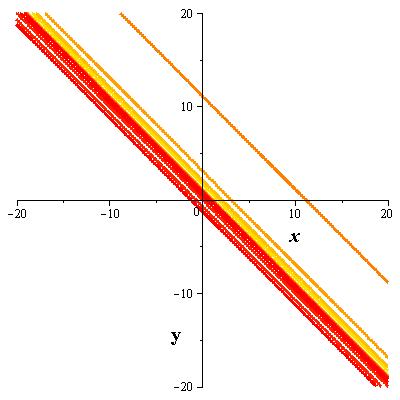}}\hspace{3cm}
    \subfigure[]{\includegraphics[width=0.32\textwidth]{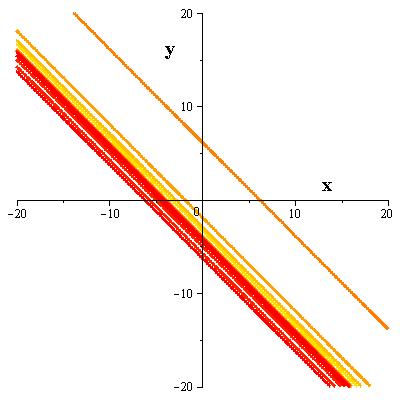}}
    \caption{Lump solutions of (\ref{example9}) at $t=0$ and $t=10$. (a), (c) $t=0$; (b), (d) $t=10$.}
    \end{figure}
\end{center}

\section{Solutions with dynamical variables}

Here we present a new kind of solutions of nonlinear partial differential equations which depend on the variables of dynamical variables. In this work we give some specific examples but study such kind of solutions later. In general we assume that the function $f$ depends on the dynamical variables $p$, $q$, $r$, etc.,  either linearly or in a nonlinear way. It is possible to connect the bilinear equations to an $n$-dimensional dynamical system, but just for illustration we take $n=3$.

\vspace{0.5cm}
\noindent
{\bf 1. Linear dynamical system}: Let $f=\beta_{0}+a_{00}\, p(\theta)+b_{00}\, q(\theta)+c_{00}\,r(\theta)$, where $p$, $q$, and $r$ depend on the variable $\theta$ satisfying
\begin{eqnarray}
&&\frac{dp}{d\theta}=a_{11}\,p+a_{12}\,q+a_{13}\,r,\label{dyna}\\
&&\frac{dq}{d\theta}=a_{21}\,p+a_{22}\,q+a_{23}\,r,\label{dynb}\\
&&\frac{dr}{d\theta}=a_{31}\,p+a_{32}\,q+a_{33}\,r,\label{dync}
\end{eqnarray}
where $a_{ij}$ ($i,j=1,2,3$) are constants and $\theta=\beta_{1}\,x+\beta_{2}\,t+\beta_{3}\,y+\beta_{4}$. When we use such an assumption into the bilinear equation (\ref{bilinearLOC1}) we obtain $10$ constraints for the $17$ constants $a_{00}, b_{00}, c_{00}, \beta_{0}, \beta_{1}, \beta_{2}, \beta_{3}, \beta_{4}$, and $a_{ij}$. The solutions of the dynamical system (\ref{dyna})-(\ref{dync}) depend on the eigenvalues of the matrix ${\bf A}={a_{ij}}$. We shall present examples of such solutions in a later communication. Next we shall give some interesting examples, the Lorentz, the Kermac-Mackendric, and the Lotka-Voltera systems where dynamical systems are nonlinear.

\vspace{0.5cm}
\noindent
{\bf 2. Lorentz system}: In this case the dynamical equation reads \cite{ay}-\cite{gurs2}
\begin{eqnarray}\label{dyn1}
&&\frac{dp}{d\theta}=\frac{q}{2},\\
&&\frac{dq}{d\theta}=-p\,r,\\
&&\frac{dr}{d\theta}=p\,q.
\end{eqnarray}
When we use this ansatz in (\ref{bilinearLOC1}) we obtain
\begin{equation}
\beta_{3}=-\frac{c_{1}}{c_{2}}\,\beta_{2},~~~\beta_{1}=\frac{\beta_{2}\,(b_{0}\,c_{1}-b_{1}\,c_{2})}{b_{1}\,c_{2}}.
\end{equation}
In addition we obtain some constraints on the system parameters: $c_{0}=0$ and
\begin{align}
-a_{0}\, b_{1}^2\, c_{2}^2-a_{1}\,b_{0}^2 c_{1}^2 +2\, a_{1}\, b_{0}\, b_{1}\,c_{1}\, c_{2}&-a_{1}\, b_{1}^2\, c_{2}^2 -a_{2}\, b_{1}^2\, c_{1}^2-a_{3}\,b_{0}\, b_{1}\, c_{1}\,c_{2}\nonumber\\
&+a_{3}\,b_{1}^2\, c_{2}^2+a_{4}\, b_{1}^2 c_{1}\, c_{2}+a_{5}\, b_{0}\, b_{1}\, c_{1}^^2-a_{5}\,b_{1}^2\, c_{1}\, c_{2}=0.
\end{align}
Hence it is possible to obtain solutions of a special case of (\ref{bilinearLOC1}) depending on the Lorentz system.

\vspace{0.5cm}
\noindent
{\bf 2. Kermac-Mackendric system}: In this case the dynamical equation reads \cite{ay}
\begin{eqnarray}\label{dyn2}
&&\frac{dp}{d\theta}=-r_{0}\,p\,q,\\
&&\frac{dq}{d\theta}=q\,(r_{0}\,p-\alpha),\\
&&\frac{dr}{d\theta}=\alpha\, q,
\end{eqnarray}
where $\alpha$ and $r_{0}$ are arbitrary constants. The only condition we obtain is $a_{00}=b_{00}=c_{00}$. This condition implies that $f=\beta_{0}+a_{00}\,(p+q+r)$. In this dynamical system $p+q+r$ is one of the conserved quantities. Hence Kermac-Mackenderic system leads to the trivial solution $f=$ constant.

\vspace{0.5cm}
\noindent
{\bf 3. Lotka-Voltera system}: This system is given as \cite{nut}
\begin{eqnarray}\label{dyn2}
&&\frac{dp}{d\theta}= p\,(\alpha\,q\,+r+\lambda),\\
&&\frac{dq}{d\theta}=q\,(p+\gamma r+\mu),\\
&&\frac{dr}{d\theta}=r\,(\rho \,p+q+\nu),
\end{eqnarray}
where $\alpha, \gamma, \rho, \mu, \nu$, and $\lambda$ are arbitrary constants. The conditions we obtain are $b_{00}=-\alpha a_{00}$, $c_{00}=-\frac{a_{00}}{\rho}$, and
\begin{align}
a_0\beta_2^2+a_1\beta_1^2&+a_2\beta_3^2+a_3\beta_1\beta_2+a_4\beta_2\beta_3+a_5\beta_1\beta_3+b_0\beta_3\beta_1^3\lambda^2\nonumber\\
&+b_1\beta_1^3\beta_2\lambda^2+b_2\beta_1^4\lambda^2+c_0\beta_1^6\lambda^4+c_1\beta_1^5\beta_2\lambda^4+c_2\beta_3\beta_1^5\lambda^4=0.
\end{align}
 In addition to these conditions  there are three restrictions for the parameters $\alpha, \gamma, \rho, \mu, \nu$, and $\lambda$,
\begin{equation}
\mu=\lambda, ~~~\nu=\lambda,~~~\rho=-\frac{1}{\gamma \alpha}.
\end{equation}
All these conditions imply that $f=\beta_{0}+a_{00}\,(p-\alpha\,q-\frac{1}{\rho}\,r)$. In this dynamical system the term $(p-\alpha\,q-\frac{1}{\rho}\,r)$ satisfies
\begin{equation}
\frac{d}{d \theta}\, (p-\alpha\,q-\frac{1}{\rho}\,r)=\lambda\, (p-\alpha\,q-\frac{1}{\rho}\,r).
\end{equation}
Hence $(p-\alpha\,q-\frac{1}{\rho}\,r)=A \, e^{\lambda \theta}$, $A$ constant, which corresponds to one-soliton solution of $f$.

The above dynamical systems are just few examples for the case of three dimensions. Some of them may lead to trivial solutions, some of which may provide the known solutions but there are examples which may give nontrivial solutions. The dynamical systems  in three dimensions are very rich \cite{ay}-\cite{gurs2}. In a later publication we shall focus on the solutions of bilinear equations as functions of nonlinear dynamical systems and on the discussion of such solutions.

\section{Conclusion}

By writing the most general Hirota bilinear form of sixth degree in $(2+1)$-dimensions we propose the most general nonlinear partial differential equation associated to this form. This differential equation covers all known Hirota integrable nonlinear partial differential equations. We have given two special cases of this differential equation: The most general local equation derivable from such a bilinear form and nonlocal differential equation derivable from this bilinear form containing the triple products. We obtained one- and two-soliton solutions of these equations and leave three-soliton solutions for later communication. The main reason for this is the constraints obtained for the existence of three-soliton solutions. We obtained also lump and hybrid solutions of our special equations. In addition we introduced a new kind of solutions of the partial differential equations connected to dynamical variables.

\section*{Appendix A: Fourth and sixth order bilinear forms}

In this Appendix we give a list of monomials Hirota bilinear forms and their associated nonlinear partial differential equations in $(2+1)$-dimensions. This list covers all monomials up to sixth order operators $D_{x}$, $D_{y}$, and $D_{t}$.
Under the transformation $u(x,y,t)=2(\ln(f(x,y,t)))_x$ we have the following equalities for the
monomials of binary products $D^{k}_{x}\,D^{2n-k}_{t}$, $D^{k}_{y}\,D^{2n-k}_{t}$, and $D^{k}_{x}\,D^{2n-k}_{y}$ for  $k=0,1,2, \cdots, 2n$, and $n=1,2,3$.

\vspace{0.5cm}
\noindent
{\bf A) Hirota bilinear forms with binary products}

\vspace{0.5cm}
\noindent
For $n=1$: All monomials are second order.
\begin{align}
&\Big(\frac{D_x^2\{f\cdot f\}}{f^2}\Big)_x=u_{xx},~~\Big(\frac{D_x\,D_{t}\{f\cdot f\}}{f^2}\Big)_x=u_{xt},\\
&\Big(\frac{D_y^2\{f\cdot f\}}{f^2}\Big)_x=u_{yy},~~\Big(\frac{D_x\, D_{y}\{f\cdot f\}}{f^2}\Big)_x=u_{xy},\\
&\Big(\frac{D_t^2\{f\cdot f\}}{f^2}\Big)_x=u_{tt},~~\Big(\frac{D_y\,D_t\{f\cdot f\}}{f^2}\Big)_x=u_{yt}.
\end{align}
For $n=2$:  All monomials are fourth order.
\begin{align}
&\Big(\frac{D_x^3D_y\{f\cdot f\}}{f^2}\Big)_x=u_{xxxy}+3u_xu_{xy}+3u_yu_{xx},\\
&\Big(\frac{D_y^3D_x\{f\cdot f\}}{f^2}\Big)_x=u_{yyyx}+3u_yu_{yy}+3u_{xy} (D^{-1} u_{yy}),\\
&\Big(\frac{D_t^3D_x\{f\cdot f\}}{f^2}\Big)_x=u_{tttx}+3u_tu_{tt}+3u_{xt} (D^{-1} u_{tt}),
\end{align}
\begin{align}
&\Big(\frac{D_t^3D_y\{f\cdot f\}}{f^2}\Big)_x=u_{ttty}+3u_{ty}(D^{-1}u_{tt})+3u_{tt}(D^{-1} u_{ty}),\\
&\Big(\frac{D_y^3D_t\{f\cdot f\}}{f^2}\Big)_x=u_{yyyt}+3u_{yt}(D^{-1} u_{yy})+3u_{yy}(D^{-1} u_{yt}),\\
&\Big(\frac{D_x^2D_t^2\{f\cdot f\}}{f^2}\Big)_x=u_{xxtt}+4u_tu_{xt}+u_xu_{tt}+u_{xx}(D^{-1}u_{tt}),\\
&\Big(\frac{D_x^2D_y^2\{f\cdot f\}}{f^2}\Big)_x=u_{xxyy}+4u_yu_{xy}+u_xu_{yy}+u_{xx}(D^{-1} u_{yy}),\\
&\Big(\frac{D_y^2D_t^2\{f\cdot f\}}{f^2}\Big)_x=u_{yytt}+u_{tt}(D^{-1} u_{yy})+u_{yy}(D^{-1} u_{tt})+4u_{yt}(D^{-1}u_{yt}),\\
&\Big(\frac{D_x^4\{f\cdot f\}}{f^2}\Big)_x=u_{xxxx}+6u_xu_{xx},\\
&\Big(\frac{D_y^4\{f\cdot f\}}{f^2}\Big)_x=u_{yyyy}+6u_{yy}(D^{-1}u_{yy}),\\
&\Big(\frac{D_t^4\{f\cdot f\}}{f^2}\Big)_x=u_{tttt}+6u_{tt}(D^{-1}u_{tt}).
\end{align}
For $n=3$:  All monomials are sixth order.
\begin{align}
&\Big(\frac{D_x^6\{f\cdot f\}}{f^2}\Big)_x=u_{6x}+15u_xu_{xxxx}+15u_{xx}u_{xxx}+45u_x^2u_{xx},\\
&\Big(\frac{D_y^6\{f\cdot f\}}{f^2}\Big)_x=u_{6y}+15u_{yyyy}(D^{-1}u_{yy})+15u_{yy}(D^{-1}u_{yyyy})+45u_{yy}( D^{-1}u_{yy})^2,\\
&\Big(\frac{D_t^6\{f\cdot f\}}{f^2}\Big)_x=u_{6t}+15u_{tttt}(D^{-1}u_{tt})+15u_{tt}(D^{-1} u_{tttt})+45u_{tt}(D^{-1} u_{tt})^2,\\
&\Big(\frac{D_x^5D_y\{f\cdot f\}}{f^2}\Big)_x=u_{xxxxxy}+10u_xu_{xxxy}+5u_{xxx}u_{xy}+5u_yu_{xxxx}+10u_{xx}u_{xxy}\nonumber\\
&\hspace{1cm}+15u_x^2u_{xy}+30u_xu_yu_{xx},\\
&\Big(\frac{D_x^5D_t\{f\cdot f\}}{f^2}\Big)_x=u_{xxxxxt}+10u_xu_{xxxt}+5u_{xxx}u_{xt}+5u_tu_{xxxx}+10u_{xx}u_{xxt}\nonumber\\
&\hspace{1cm}+15u_x^2u_{xt}+30u_xu_tu_{xx},\\
&\Big(\frac{D_x^4D_y^2\{f\cdot f\}}{f^2}\Big)_x=u_{xxxxyy}+6u_xu_{xxyy}+6u_{xx}u_{xyy}+8u_{xy}u_{xxy}+8u_{y}u_{xxxy}\nonumber\\
&\hspace{1cm}+u_{xxx}u_{yy}+u_{xxxx}(D^{-1}u_{yy}),\\
&\Big(\frac{D_x^4D_t^2\{f\cdot f\}}{f^2}\Big)_x=u_{xxxxtt}+6u_xu_{xxtt}+6u_{xx}u_{xtt}+8u_{xt}u_{xxt}+8u_{t}u_{xxxt}\nonumber\\
&\hspace{1cm}+u_{xxx}u_{tt}+u_{xxxx}(D^{-1}u_{tt}),\\
&\Big(\frac{D_x^3D_t^3\{f\cdot f\}}{f^2}\Big)_x=u_{xxxttt}+9u_tu_{xxtt}+3u_{tt}u_{xxt}+3u_{xx}u_{ttt}+9u_{t}u_xu_{tt}\nonumber\\
&\hspace{1cm}+9u_{xt}u_{xtt}+3u_{x}u_{xttt}+18u_t^2u_{xt}+\Big[3u_{xxxt}+9u_tu_{xx}+9u_xu_{xt}\Big](D^{-1}u_{tt}),
\end{align}
\begin{align}
&\Big(\frac{D_x^3D_y^3\{f\cdot f\}}{f^2}\Big)_x=u_{xxxyyy}+9u_yu_{xxyy}+3u_{yy}u_{xxy}+3u_{xx}u_{yyy}+9u_{y}u_xu_{yy}\nonumber\\
&\hspace{1cm}+9u_{xy}u_{xyy}+3u_{x}u_{xyyy}+18u_y^2u_{xy}+\Big[3u_{xxxy}+9u_yu_{xx}+9u_xu_{xy}\Big](D^{-1}u_{yy}),\\
&\nonumber\\
&\Big(\frac{D_x^2D_t^4\{f\cdot f\}}{f^2}\Big)_x=u_{xxtttt}+8u_tu_{xttt}+6u_{tt}u_{xtt}+8u_{xt}u_{ttt}+12u_t^2u_{tt}+u_{xx}(D^{-1}u_{tttt})\nonumber\\
&+u_xu_{tttt}+\Big[6u_{xxtt}+24u_tu_{xt}+6u_xu_{tt}+3u_{xx}(D^{-1}u_{tt})\Big](D^{-1}u_{tt}),\\
&\nonumber\\
&\Big(\frac{D_x^2D_y^4\{f\cdot f\}}{f^2}\Big)_x=u_{xxyyyy}+8u_yu_{xyyy}+6u_{yy}u_{xyy}+8u_{xy}u_{yyy}+12u_y^2u_{yy}+u_{xx}(D^{-1}u_{yyyy})\nonumber\\
&+u_xu_{yyyy}+\Big[6u_{xxyy}+24u_yu_{xy}+6u_xu_{yy}+3u_{xx}(D^{-1}u_{yy})\Big](D^{-1}u_{yy}),\\
&\nonumber\\
&\Big(\frac{D_y^5D_x\{f\cdot f\}}{f^2}\Big)_x=u_{xyyyyy}+5u_yu_{yyyy}+10u_{yy}u_{yyy}+5u_{xy}(D^{-1}u_{yyyy})\nonumber\\
&+\Big[10u_{xyyy}+30u_yu_{yy}+15u_{xy}(D^{-1}u_{yy})\Big](D^{-1}u_{yy}),\\
&\nonumber\\
&\Big(\frac{D_t^5D_x\{f\cdot f\}}{f^2}\Big)_x=u_{xttttt}+5u_tu_{tttt}+10u_{tt}u_{ttt}+5u_{xt}(D^{-1}u_{tttt})\nonumber\\
&+\Big[10u_{xttt}+30u_yu_{tt}+15u_{xt}(D^{-1}u_{tt})\Big](D^{-1}u_{tt}),\\
&\nonumber\\
&\Big(\frac{D_y^5D_t\{f\cdot f\}}{f^2}\Big)_x=u_{yyyyyt}+10u_{yy}(D^{-1}u_{yyyt})+10u_{yyyt}(D^{-1}u_{yy})+5u_{yt}(D^{-1}u_{yyyy})\nonumber\\
&+5u_{yyyy}(D^{-1}u_{yt})+15u_{yt}(D^{-1}u_{yy})^2+30u_{yy}(D^{-1}u_{yy})(D^{-1}u_{yt}),\\
&\nonumber\\
&\Big(\frac{D_t^5D_y\{f\cdot f\}}{f^2}\Big)_x=u_{ttttty}+10u_{tt}(D^{-1}u_{ttty})+10u_{ttty}(D^{-1}u_{tt})+5u_{yt}(D^{-1}u_{tttt})\nonumber\\
&+5u_{tttt}(D^{-1} u_{yt})+15u_{yt}(D^{-1}u_{tt})^2+30u_{tt}(D^{-1}u_{tt})(D^{-1}u_{yt}),\\
&\nonumber\\
&\Big(\frac{D_y^4D_t^2\{f\cdot f\}}{f^2}\Big)_x=u_{yyyytt}+u_{tt}(D^{-1} u_{yyyy})+u_{yyyy}(D^{-1}u_{tt})+6u_{yy}(D^{-1}u_{tt})(D^{-1}u_{yy})
\nonumber\\
&+24u_{yt}(D^{-1}u_{yt})(D^{-1}u_{yy})
+6u_{yy}(D^{-1}u_{yytt})
+8u_{yt}(D^{-1}u_{yyyt})\nonumber\\
&+8u_{yyyt}(D^{-1}u_{yt})
+6u_{yytt}(D^{-1}u_{yy})+3u_{tt}(D^{-1}u_{yy})^2+2u_{yy}(D^{-1}u_{yt})^2,\\
&\nonumber\\
&\Big(\frac{D_t^4D_y^2\{f\cdot f\}}{f^2}\Big)_x=u_{ttttyy}+u_{yy}(D^{-1} u_{tttt})+u_{tttt}(D^{-1}u_{yy})
+6u_{tt}(D^{-1}u_{yy})(D^{-1}u_{tt})
\nonumber\\
&+24u_{yt}(D^{-1}u_{yt})(D^{-1}u_{tt})
+6u_{tt}(D^{-1}u_{yytt})
+8u_{yt}(D^{-1}u_{ttty})\nonumber\\
&+8u_{ttty}(D^{-1}u_{yt})
+6u_{yytt}(D^{-1}u_{tt})
+3u_{yy}(D^{-1}u_{tt})^2+2u_{tt}(D^{-1}u_{yt})^2,\\
&\nonumber
\end{align}
\begin{align}
&\Big(\frac{D_y^3D_t^3\{f\cdot f\}}{f^2}\Big)_x=u_{yyyttt}+3u_{yttt}(D^{-1}u_{yy})+9u_{yt}(D^{-1}u_{yytt})+9u_{yytt}(D^{-1}u_{yt})\nonumber\\
&+3u_{tt}(D^{-1}u_{yyyt})+3u_{yyyt}(D^{-1}u_{tt})+18u_{yt}(D^{-1}u_{yt})^2+3u_{yy}(D^{-1}u_{yttt})\nonumber\\
&+9u_{yt}(D^{-1}u_{tt})(D^{-1}u_{yy})+9u_{yy}(D^{-1}u_{yt})(D^{-1}u_{tt})
\nonumber\\&+9u_{tt}(D^{-1}u_{yt})(D^{-1}u_{yy}).
\end{align}

\noindent
{\bf B) Hirota bilinear forms with triple products}

\vspace{0.5cm}
\noindent
For $n=2$: Hirota bilinear forms of fourth order.
\begin{align}
&\Big(\frac{D_{t}\,D_x^2\,D_y \,\{f\cdot f\}}{f^2}\Big)_x=u_{xxyt}+u_xu_{yt}+2u_yu_{xt}+2u_tu_{xy}+u_{xx}(D^{-1}u_{yt}),\\
&\Big(\frac{D_{t}^2\,D_x\,D_y \,\{f\cdot f\}}{f^2}\Big)_x=u_{xytt}+u_yu_{tt}+2u_tu_{yt}+u_{xy}(D^{-1}u_{tt})+2u_{xt}(D^{-1}u_{yt}),\\
&\Big(\frac{D_{t}\,D_x\,D_y^2 \,\{f\cdot f\}}{f^2}\Big)_x=u_{xtyy}+u_tu_{yy}+2u_yu_{yt}+u_{xt}(D^{-1}u_{yy})+2u_{xy}(D^{-1}u_{yt}).
\end{align}
\vspace{0.4cm}
\noindent
For $n=3$: Hirota bilinear forms of order six.
\begin{align}
&\Big(\frac{D_{t}\,D_x\,D_y^4 \,\{f\cdot f\}}{f^2}\Big)_x=u_{yyyyxt}+u_tu_{yyyy}+4u_{yt}u_{yyy}+6u_{yy}u_{tyy}+4u_yu_{tyyy}+u_{xt}(D^{-1}u_{yyyy})\nonumber\\
&+\Big(6u_tu_{yy}+12u_yu_{yt}+6u_{xtyy}+3u_{xt}(D^{-1}u_{yy})\Big)(D^{-1}u_{yy})+4u_{xy}(D^{-1}u_{tyyy})\nonumber\\
&+\Big(4u_{xyyy}+12u_yu_{yy}+12u_{xy}(D^{-1}u_{yy})\Big)(D^{-1}u_{yt}),\\
&\nonumber\\
&\Big(\frac{D_{t}\,D_x^2\,D_y^3 \,\{f\cdot f\}}{f^2}\Big)_x=u_{yyyxxt}+u_xu_{yyyt}+6u_yu_{xyyt}+6u_y^2u_{yt}+2u_tu_{xyyy}+6u_tu_yu_{yy}\nonumber\\
&+6u_{xy}u_{yyt}+3u_{yt}u_{xyy}+2u_{yyy}u_{xt}+\Big(3u_xu_{yt}+6u_yu_{xt}+6u_tu_{xy}+3u_{xxyt}\Big)(D^{-1}u_{yy})\nonumber\\
&+3u_{yy}u_{xyt}+\Big(3u_{xxyy}+12u_yu_{xy}+3u_xu_{yy}+3u_{xx}(D^{-1}u_{yy})\Big)(D^{-1}u_{yt})+u_{xx}(D^{-1}u_{yyyt}),\\
&\nonumber\\
&\Big(\frac{D_{t}\,D_x^3\,D_y^2 \,\{f\cdot f\}}{f^2}\Big)_x=u_{xxxyyt}+6u_{xy}u_{xyt}+3u_{xx}u_{yyt}+2u_{yt}u_{xxy}+3u_{xt}u_{xyy}+6u_yu_{xxyt}\nonumber\\
&+3u_xu_{xyyt}+6u_{xt}u_y^2+3u_tu_{xxyy}+3u_xu_tu_{yy}+u_{yy}u_{xxt}+12u_tu_yu_{xy}+6u_xu_yu_{yt}\nonumber\\
&+\Big(2u_{xxxy}+6u_xu_{xy}+6u_yu_{xx}\Big)(D^{-1}u_{yt})+\Big(u_{xxxt}+3u_xu_{xt}+3u_tu_{xx}\Big)(D^{-1}u_{yy}),\\
&\nonumber
\end{align}
\begin{align}
&\Big(\frac{D_{t}\,D_x^4\,D_y \,\{f\cdot f\}}{f^2}\Big)_x=u_{xxxxyt}+12u_tu_yu_{xx}+u_{xxx}u_{yt}+12u_xu_yu_{xt}+12u_xu_tu_{xy}\nonumber\\
&+4u_{xy}u_{xxt}+6u_{xx}u_{xyt}+4u_{xt}u_{xxy}+4u_yu_{xxxt}+6u_xu_{xxyt}+3u_x^2u_{yt}+4u_tu_{xxxy}\nonumber\\
&+\Big(u_{xxxx}+6u_xu_{xx}\Big)(D^{-1}u_{yt}),\\
&\nonumber\\
&\Big(\frac{D_{x}\,D_t^2\,D_y^3 \,\{f\cdot f\}}{f^2}\Big)_x=u_{yyyxtt}+3u_{ytt}u_{yy}+6u_{yt}u_{yyt}+3u_yu_{yytt}+2u_tu_{yyyt}+u_{tt}u_{yyy}\nonumber\\
&+\Big(3u_{xytt}+3u_{y}u_{tt}+6u_tu_{yt}+6u_{xt}(D^{-1}u_{yt})\Big)(D^{-1}u_{yy})+3u_{xy}(D^{-1}u_{yytt})\nonumber\\
&+\Big(u_{xyyy}+3u_yu_{yy}+3u_{xy}(D^{-1}u_{yy})\Big)(D^{-1}u_{tt})+2u_{xt}(D^{-1}u_{yyyt})\nonumber\\
&+\Big(6u_{xyyt}+6u_tu_{yy}+12u_yu_{yt}+6u_{xy}(D^{-1}u_{yt})\Big)(D^{-1}u_{yt}),\\
&\nonumber\\
&\Big(\frac{D_{x}\,D_t^4\,D_y \,\{f\cdot f\}}{f^2}\Big)_x=u_{ttttxy}+u_yu_{tttt}+4u_{yt}u_{ttt}+6u_{tt}u_{ytt}+4u_tu_{yttt}+u_{xy}(D^{-1}u_{tttt})\nonumber\\
&+\Big(6u_yu_{tt}+12u_tu_{yt}+6u_{xytt}+3u_{xy}(D^{-1}u_{tt})\Big)(D^{-1}u_{tt})+4u_{xt}(D^{-1}u_{yttt})\nonumber\\
&+\Big(4u_{xttt}+12u_tu_{tt}+12u_{xt}(D^{-1}u_{tt})\Big)(D^{-1}u_{yt}),\\
&\nonumber\\
&\Big(\frac{D_{x}\,D_t^3\,D_y^2 \,\{f\cdot f\}}{f^2}\Big)_x=u_{tttxyy}+3u_{tyy}u_{tt}+6u_{yt}u_{tty}+3u_tu_{yytt}+2u_yu_{ttty}+u_{yy}u_{ttt}\nonumber\\
&+\Big(3u_{xtyy}+3u_{t}u_{yy}+6u_yu_{yt}+6u_{xy}(D^{-1}u_{yt})\Big)(D^{-1}u_{tt})+3u_{xt}(D^{-1}u_{yytt})\nonumber\\
&+\Big(u_{xttt}+3u_tu_{tt}+3u_{xt}(D^{-1}u_{tt})\Big)(D^{-1}u_{yy})+2u_{xy}(D^{-1}u_{ttty})\nonumber\\
&+\Big(6u_{xtty}+6u_yu_{tt}+12u_tu_{yt}+6u_{xt}(D^{-1}u_{yt})\Big)(D^{-1}u_{yt}),\\
&\nonumber\\
&\Big(\frac{D_{y}\,D_x^2\,D_t^3 \,\{f\cdot f\}}{f^2}\Big)_x=u_{tttxxy}+u_xu_{ttty}+6u_tu_{xtty}+6u_t^2u_{yt}+2u_yu_{xttt}+6u_tu_yu_{tt}+3u_{tt}u_{xyt}\nonumber\\
&+6u_{xt}u_{tty}+3u_{yt}u_{xtt}+2u_{ttt}u_{xy}+\Big(3u_xu_{yt}+6u_tu_{xy}+3u_{xxyt}+6u_yu_{xt}\Big)(D^{-1}u_{tt})\nonumber\\
&+\Big(3u_{xxtt}+3u_xu_{tt}+12u_tu_{xt}+3u_{xx}(D^{-1}u_{tt})\Big)(D^{-1}u_{yt})+u_{xx}(D^{-1}u_{ttty}),\\
&\nonumber\\
&\Big(\frac{D_{y}\,D_x^3\,D_t^2 \,\{f\cdot f\}}{f^2}\Big)_x=u_{xxxtty}+6u_{xt}u_{xyt}+3u_{xx}u_{tty}+2u_{yt}u_{xxt}+3u_{xy}u_{xtt}+6u_tu_{xxyt}\nonumber\\
&+3u_xu_{xtty}+6u_{xy}u_t^2+3u_yu_{xxtt}+3u_xu_yu_{tt}+u_{tt}u_{xxy}+12u_tu_yu_{xt}+6u_xu_tu_{yt}\nonumber\\
&+\Big(2u_{xxxt}+6u_xu_{xt}+6u_tu_{xx}\Big)(D^{-1}u_{yt})+\Big(u_{xxxy}+3u_xu_{xy}+3u_yu_{xx}\Big)(D^{-1}u_{tt}),
\end{align}
\begin{align}
&\Big(\frac{D_{x}^2\,D_t^2\,D_y^2 \,\{f\cdot f\}}{f^2}\Big)_x=u_{xxyytt}+8u_tu_yu_{yt}+u_{yy}u_{xtt}+u_xu_{yytt}+u_{tt}u_{yyx}+4u_{xt}u_{yyt}
\nonumber\\
&+4u_yu_{xytt}+4u_tu_{yyxt}+4u_{yx}u_{ytt}+2u_y^2u_{tt}+2u_t^2u_{yy}+4u_{yt}u_{xyt}\nonumber\\
&+u_{xx}(D^{-1}u_{yytt})+\Big(4u_yu_{xy}+u_xu_{yy}+u_{yyxx}\Big)(D^{-1}u_{tt})\nonumber\\
&+\Big(4u_tu_{xt}+u_xu_{tt}+u_{xxtt}+u_{xx}(D^{-1}u_{tt})\Big)(D^{-1}u_{yy})\nonumber\\
&+\Big(8u_yu_{xt}+4u_xu_{yt}+8u_tu_{xy}+4u_{yxxt}+2u_{xx}(D^{-1} u_{yt})\Big)(D^{-1}u_{yt}).
\end{align}

\section*{Appendix B:  Conditions for the lump solutions}

Lump solutions of differential equations are obtained by letting
\begin{equation}\label{3lumpgen}
f=\beta_0+p^2+q^2+r^2,
\end{equation}
where  $p=\beta_1 x+\beta_2 y+\beta_3 t+\beta_4$, $q=\beta_5 x+\beta_6 y+\beta_7 t+\beta_8$, and $r=\beta_9 x+\beta_{10} y+\beta_{11} t+\beta_{12}$  where $\beta_{0}, \beta_{1}, \cdots, \beta_{12}$ are arbitrary constants.
The system of equations satisfied by $\beta_{j}$'s obtained from the bilinear equations (\ref{bilinearLOC1}) and (\ref{bilinearNON1}) are given below.

\vspace{0.5cm}
\noindent
{\bf 1. For the equation (\ref{bilinearLOC1})}
\begin{align}
&1)\, 6(\beta_1^2+\beta_5^2+\beta_9^2)^2b_2+6(\beta_1^2+\beta_5^2+\beta_9^2)(\beta_1\beta_3+\beta_1\beta_9+\beta_5\beta_7)b_1\nonumber\\
&+6(\beta_1^2+\beta_5^2+\beta_9^2)(\beta_1\beta_2+\beta_9\beta_{10}+\beta_5\beta_6)b_0+\beta_0(\beta_3^2+\beta_7^2+\beta_{11}^2)a_0\nonumber\\
&+\beta_0(\beta_1^2+\beta_5^2+\beta_9^2)a_1+\beta_0(\beta_2^2+\beta_6^2+\beta_{10}^2)a_2+\beta_0(\beta_1\beta_3+\beta_9\beta_{11}+\beta_5\beta_7)a_3\nonumber\\
&+\beta_0(\beta_6\beta_7+\beta_{10}\beta_{11}+\beta_2\beta_3)a_4+\beta_0(\beta_1\beta_2+\beta_9\beta_{10}+\beta_5\beta_6)a_5=0,\label{app3luloc1}\\
&2)\,(\beta_2\beta_5+\beta_1\beta_6)a_5+(\beta_3\beta_6+\beta_2\beta_7)a_4+(\beta_3\beta_5+\beta_1\beta_7)a_3+2\beta_2\beta_6a_2\nonumber\\
&+2\beta_1\beta_5a_1+2\beta_3\beta_7a_0=0,\label{app3luloc2}\\
&3)\,(\beta_6\beta_9+\beta_5\beta_{10})a_5+(\beta_6\beta_{11}+\beta_7\beta_{10})a_4+(\beta_5\beta_{11}+\beta_7\beta_{9})a_3+2\beta_6\beta_{10}a_2\nonumber\\
&+2\beta_5\beta_9a_1+2\beta_7\beta_{11}a_0=0,\label{app3luloc3}\\
&4)\,(\beta_1\beta_{10}+\beta_2\beta_9)a_5+(\beta_3\beta_{10}+\beta_2\beta_{11})a_4+(\beta_1\beta_{11}+\beta_3\beta_9)a_3+2\beta_2\beta_{10}a_2\nonumber\\
&+2\beta_1\beta_9a_1+2\beta_3\beta_{11}a_0=0,\label{app3luloc4}\\
&5)\,(\beta_9\beta_{10}+\beta_1\beta_2-\beta_5\beta_6)a_5+(\beta_2\beta_3+\beta_{10}\beta_{11}-\beta_6\beta_7)a_4+(\beta_1\beta_3+\beta_{9}\beta_{11}-\beta_5\beta_7)a_3\nonumber\\
&+(\beta_{10}^2+\beta_2^2-\beta_6^2)a_2+(\beta_9^2+\beta_1^2-\beta_5^2)a_1+(\beta_{11}^2+\beta_3^2-\beta_7^2)a_0=0,\label{app3luloc5}\\
&6)\,(\beta_1\beta_{2}+\beta_5\beta_6-\beta_9\beta_{10})a_5+(\beta_2\beta_3+\beta_{6}\beta_{7}-\beta_{10}\beta_{11})a_4+(\beta_5\beta_7+\beta_{1}\beta_{3}-\beta_9\beta_{11})a_3\nonumber\\
&+(\beta_{6}^2+\beta_2^2-\beta_{10}^2)a_2+(\beta_5^2+\beta_1^2-\beta_9^2)a_1+(\beta_{7}^2+\beta_3^2-\beta_{11}^2)a_0=0,\label{app3luloc6}\\
&7)\,(\beta_5\beta_{6}+\beta_9\beta_{10}-\beta_1\beta_2)a_5+(\beta_6\beta_7+\beta_{10}\beta_{11}-\beta_2\beta_3)a_4+(\beta_9\beta_{11}+\beta_{5}\beta_{7}-\beta_1\beta_3)a_3\nonumber\\
&+(\beta_{10}^2+\beta_6^2-\beta_2^2)a_2+(\beta_9^2+\beta_5^2-\beta_1^2)a_1+(\beta_{7}^2+\beta_{11}^2-\beta_3^2)a_0=0.\label{app3luloc7}
\end{align}

From the last three equations we get
{\small
\begin{align}
&\beta_1=\frac{1}{2a_1}[-a_3\beta_3-a_5\beta_2+\sqrt{a_5^2\beta_2^2+2a_5a_3\beta_2\beta_3+a_3^2\beta_3^2-4a_1a_2\beta_2^2-4a_1a_4\beta_2\beta_3-4a_0a_1\beta_3^2}],\\
&\beta_5=\frac{1}{2a_1}[-a_3\beta_7-a_5\beta_6+\sqrt{a_3^2\beta_7^2+2a_5a_3\beta_6\beta_7+a_5^2\beta_6^2-4a_1a_2\beta_6^2-4a_1a_4\beta_6\beta_7-4a_0a_1\beta_7^2}],\\
&\beta_9=\frac{1}{2a_1}[-a_3\beta_{11}-a_5\beta_{10}\nonumber\\
&\hspace{1.5cm}+\sqrt{a_3^2\beta_{11}^2+2a_5a_3\beta_{10}\beta_{11}+a_5^2\beta_{10}^2-4a_1a_2\beta_{10}^2-4a_1a_4\beta_{10}\beta_{11}
-4a_0a_1\beta_{11}^2}].
\end{align}}
Using the above equalities in the equations (\ref{app3luloc2})-(\ref{app3luloc4}) gives
\begin{equation}\displaystyle
\beta_2=\frac{\beta_3\beta_{10}}{\beta_{11}},\quad \beta_6=\frac{\beta_7\beta_{10}}{\beta_{11}},\label{3lulocrel1}
\end{equation}
yielding
\begin{equation}\label{3lulocrel2}
\beta_1=\eta \beta_3,\quad \beta_5=\eta \beta_7,\quad \beta_9=\eta \beta_{11},
\end{equation}
where
\begin{align}\label{3luloceta}\displaystyle
\eta=&\frac{1}{2a_1\beta_{11}}\sqrt{a_5^2\beta_{10}^2+2a_5a_3\beta_{10}\beta_{11}+a_3^2\beta_{11}^2-4a_1a_2\beta_{10}^2-4a_1a_4\beta_{10}\beta_{11}-4a_0a_1\beta_{11}^2}\nonumber\\
&-\frac{a_5\beta_{10}}{2a_1\beta_{11}}-\frac{a_3}{2a_1}.
\end{align}
Under these results the equation (\ref{app3luloc1}) turns to be
\begin{align}
&6b_2(\beta_3^2+\beta_7^2+\beta_{11}^2)\eta^4+\frac{6(\beta_3^2+\beta_7^2+\beta_{11}^2)}{\beta_{11}}(b_0\beta_{10}+b_1\beta_{11})\eta^3+a_1\beta_0\eta^2\nonumber\\
&+\beta_0(a_3\beta_{11}+a_5\beta_{10})\eta+\frac{\beta_0}{\beta_{11}^2}(a_2\beta_{10}^2+a_4\beta_{10}\beta_{11}+a_0\beta_{11}^2)=0.
\end{align}
\vspace{0.4cm}
\noindent
{\bf 2. For the equation  (\ref{bilinearNON1}) }
\begin{align}
&1)\,(\beta_6^2+\beta_2^2-\beta_{10}^2)a_1+(\beta_1\beta_3+\beta_5\beta_7-\beta_9\beta_{11})a_0=0, \label{uclunon-1}\\
&2)\, (\beta_{10}^2+\beta_2^2-\beta_{6}^2)a_1+(\beta_1\beta_3+\beta_9\beta_{11}-\beta_5\beta_{7})a_0=0,\label{uclunon-2}\\
&3)\, (\beta_6^2+\beta_{10}^2-\beta_{2}^2)a_1+(\beta_5\beta_7+\beta_9\beta_{11}-\beta_1\beta_{3})a_0=0,\label{uclunon-3}\\
&4)\, (\beta_3\beta_5+\beta_1\beta_7)a_0+2\beta_2\beta_6a_1=0,\label{uclunon-4}\\
&5)\, (\beta_1\beta_{11}+\beta_3\beta_9)a_0+2\beta_2\beta_{10}a_1=0,\label{uclunon-5}\\
&6)\, (\beta_5\beta_{11}+\beta_7\beta_9)a_0+2\beta_6\beta_{10}a_1=0,\label{uclunon-6}
\end{align}
\begin{align}
&7)\,2\Big(\beta_2^2[3\beta_1\beta_3+\beta_5\beta_7+\beta_9\beta_{11}]+\beta_6^2[\beta_1\beta_3+3\beta_5\beta_7+\beta_9\beta_{11}]\nonumber\\
&+\beta_{10}^2[\beta_1\beta_3+\beta_5\beta_7+3\beta_9\beta_{11}]+2\beta_1\beta_2[\beta_6\beta_7+\beta_{10}\beta_{11}]+2\beta_5\beta_6[\beta_2\beta_3+\beta_{10}\beta_{11}]\nonumber\\
&+2\beta_9\beta_{10}[\beta_2\beta_3+\beta_6\beta_7]\Big)b_2+2\Big(\beta_3^2[3\beta_1\beta_2+\beta_5\beta_6+\beta_9\beta_{10}]+\beta_7^2[\beta_1\beta_2+3\beta_5\beta_6+\beta_9\beta_{10}]\nonumber\\
&+\beta_{11}^2[\beta_1\beta_2+\beta_5\beta_6+3\beta_9\beta_{10}]+2\beta_1\beta_3[\beta_6\beta_7+\beta_{10}\beta_{11}]+2\beta_2\beta_3[\beta_6\beta_7+\beta_{9}\beta_{11}]\nonumber\\
&+2\beta_5\beta_{7}[\beta_2\beta_3+\beta_{10}\beta_{11}]\Big)b_1+2\Big(\beta_5^2[3\beta_6\beta_7+\beta_2\beta_3+\beta_{10}\beta_{11}]+\beta_1^2[\beta_6\beta_7+3\beta_2\beta_3+\beta_{10}\beta_{11}]\nonumber\\
&+\beta_{9}^2[\beta_6\beta_7+\beta_2\beta_3+3\beta_{10}\beta_{11}]+2\beta_5\beta_7[\beta_1\beta_2+\beta_{9}\beta_{10}]+2\beta_5\beta_6[\beta_1\beta_3+\beta_{9}\beta_{11}]\nonumber\\
&+2\beta_1\beta_{9}[\beta_2\beta_{11}+\beta_3\beta_{10}]\Big)b_0+6(\beta_1^2+\beta_5^2+\beta_9^2)^2a_2+\beta_0(\beta_2^2+\beta_6^2+\beta_{10}^2)a_1\nonumber\\
&+\beta_0(\beta_9\beta_{11}+\beta_1\beta_3+\beta_5\beta_7)a_0=0.\label{uclunon-7}
\end{align}

From the first three equations we obtain
\begin{eqnarray}
&&\beta_{2}^2\, a_{1}+\beta_{1}\, \beta_{3} a_{0}=0, \\
&&\beta_{6}^2 \,a_{1}+\beta_{5}\, \beta_{7} a_{0}=0, \\
&&\beta_{10}^2 \,a_{1}+\beta_{11}\, \beta_{9} a_{0}=0.
\end{eqnarray}
Hence
\begin{equation}
\beta_{1}=-\frac{\beta_{2}^2}{\beta_{3}} \lambda, ~~\beta_{5}=-\frac{\beta_{6}^2}{\beta_{7}} \lambda,~~\beta_{9}=-\frac{\beta_{10}^2}{\beta_{7}} \lambda,
\end{equation}
where we assume $\beta_{3}, \beta_{7}, \beta_{11}$, and $a_{0}$ are different than zero. Here $\lambda=\frac{a_{1}}{a_{0}}$. By using the above $\beta_j$'s, the next three equations (\ref{uclunon-4})-(\ref{uclunon-6}) give
\begin{equation}
\beta_{2}=\frac{\beta_{3}\, \beta_{10}}{\beta_{11}},~~~~\beta_{6}=\frac{\beta_{7}\, \beta_{10}}{\beta_{11}}.
\end{equation}
Inserting all the equalities that we obtain for $\beta_j$'s in the last equation we obtain a single equation for $\beta_{10}$, $\beta_{11}$, and $\lambda$ as
\begin{equation}
\lambda b_0\beta_{10}^2\beta_{11}^3+\lambda^3a_2\beta_{10}^5-b_2\beta_{10}\beta_{11}^4-b_1\beta_{11}^5=0.
\end{equation}
Letting $\mu=\frac{\beta_{10}}{\beta_{11}}$, then $\mu$ satisfies the following nonlinear algebraic equation
\begin{equation}
\lambda b_0 \, \mu^2+\lambda^3a_2\, \mu^5-b_2\, \mu-b_1=0.
\end{equation}
\newpage

\section*{Appendix C: Conditions for the hybrid solutions}

\noindent \textbf{Case 1}

\noindent \textbf{1. For the equation (\ref{bilinearLOC1})}
\begin{align}
& 1)\, \beta_1(b_0\beta_2+b_1\beta_3+b_2\beta_1)=0,\label{case1loc1}\\
& 2)\, a_0\beta_3^2+a_1\beta_1^2+a_2\beta_2^2+a_3\beta_1\beta_3+a_4\beta_2\beta_3+a_5\beta_1\beta_2=0,\label{case1loc2}\\
& 3)\, a_0\beta_7^2+a_1\beta_5^2+a_2\beta_6^2+a_3\beta_5\beta_7+a_4\beta_6\beta_7+a_5\beta_5\beta_6+\beta_5^3(b_0\beta_6+b_1\beta_7+b_2\beta_5)
\nonumber\\
&+\beta_5^5(c_0\beta_5+c_1\beta_7+c_2\beta_6)=0,\label{case1loc3}\\
& 4)\, \beta_1\beta_5[3(\beta_2\beta_5+\beta_1\beta_6)b_0+3(\beta_3\beta_5+\beta_1\beta_7)b_1+6\beta_1\beta_5b_2+15\beta_1\beta_5^3c_0\nonumber\\
&+5\beta_5^2(\beta_3\beta_5+2\beta_1\beta_7)c_1+5\beta_5^2(\beta_2\beta_5+2\beta_1\beta_6)c_2]=0,\label{case1loc4}\\
& 5)\, 2a_0\beta_3\beta_7+2a_1\beta_1\beta_5+2a_2\beta_2\beta_6+a_3(\beta_1\beta_7+\beta_3\beta_5)+a_4(\beta_2\beta_7+\beta_3\beta_6)+a_5(\beta_1\beta_6+\beta_2\beta_5)
\nonumber\\&+b_0(\beta_2\beta_5^3+3\beta_5^2\beta_1\beta_6)+b_1(\beta_3\beta_5^3+3\beta_1\beta_7\beta_5^2)+4b_2\beta_1\beta_5^3+6c_0\beta_1\beta_5^5
+c_1(\beta_3\beta_5^5+5\beta_1\beta_7\beta_5^4)\nonumber\\
&+c_2(\beta_2\beta_5^5+5\beta_1\beta_6\beta_5^4)=0.\label{case1loc5}
\end{align}

\noindent \textbf{2. For the equation (\ref{bilinearNON1})}
\begin{align}
&1)\, a_0\beta_1\beta_3+a_1\beta_2^2=0,\label{case1noneq1}\\
&2)\, \beta_1(a_1\beta_1^3+b_0\beta_1\beta_2\beta_3+b_1\beta_2\beta_3^2+b_2\beta_2^2\beta_3)=0,\label{case1noneq2}\\
&3)\, a_0\beta_5\beta_7+a_1\beta_6^2+a_2\beta_5^4+b_0\beta_5^2\beta_6\beta_7+b_1\beta_5\beta_6\beta_7^2+b_2\beta_5\beta_6^2\beta_7=0,\label{case1noneq3}\\
&4)\, 6a_2\beta_1^2\beta_5^2+(\beta_2\beta_3\beta_5^2+2\beta_1\beta_2\beta_5\beta_7+\beta_6\beta_7\beta_1^2+2\beta_1\beta_3\beta_5\beta_6)b_0\nonumber\\
&+(\beta_1\beta_2\beta_7^2+2\beta_1\beta_3\beta_6\beta_7+\beta_5\beta_6\beta_3^2+2\beta_2\beta_3\beta_5\beta_7)b_1\nonumber\\
&+(\beta_5\beta_7\beta_2^2+2\beta_1\beta_2\beta_6\beta_7+\beta_1\beta_3\beta_6^2+2\beta_2\beta_3\beta_5\beta_6)b_1=0,\label{case1noneq4}\\
&5)\, (\beta_3\beta_5+\beta_1\beta_7)a_0+2a_1\beta_2\beta_6+4a_2\beta_1\beta_5^3+(\beta_5^2\beta_3\beta_6+2\beta_1\beta_5\beta_6\beta_7+\beta_5^2\beta_2\beta_7)b_0\nonumber\\
&+(\beta_7^2\beta_1\beta_6+2\beta_3\beta_5\beta_6\beta_7+\beta_7^2\beta_2\beta_5)b_1+(\beta_6^2\beta_3\beta_5+2\beta_2\beta_5\beta_6\beta_7
+\beta_6^2\beta_1\beta_7)b_2=0.\label{case1noneq5}
\end{align}
\noindent \textbf{Case 2}

\noindent In this case the conditions (\ref{case1loc1})-(\ref{case1loc4}) given for Case 1 are same with the conditions obtained for the equation (\ref{bilinearLOC1}).
The fifth constraint for (\ref{bilinearLOC1}) is the condition (\ref{case1loc5}) multiplied by $(\beta_0A-1)$. There is also sixth constraint given below:
\begin{equation}
(1+\beta_0A)\varphi+60A\beta_1^3\beta_5(3\beta_1\beta_5c_0+(\beta_1\beta_7+2\beta_3\beta_5)c_1+(\beta_1\beta_6+2\beta_2\beta_5)c_2)=0,
\end{equation}
where $\varphi$ is the condition (\ref{case1loc4}).

\noindent For the equation (\ref{bilinearNON1}), the conditions (\ref{case1noneq1})-(\ref{case1noneq4}) given for Case 1 are same with the ones obtained for Case 2.
The fifth constraint for (\ref{bilinearNON1}) is the condition (\ref{case1noneq5}) multiplied by $(\beta_0A-1)$.

%\section{Acknowledgment}
%  This work is partially supported by the Scientific
%and Technological Research Council of Turkey (T\"{U}B\.{I}TAK).\\

\end{document}